\documentclass[english,reprint]{revtex4-1}
\usepackage[LGR,T1]{fontenc}
\usepackage[latin9]{inputenc}
\usepackage{color}
\usepackage{babel}
\usepackage{amsmath}
\usepackage{graphicx}
\usepackage{esint}
\usepackage{subscript}
\usepackage[unicode=true,pdfusetitle,
 bookmarks=true,bookmarksnumbered=false,bookmarksopen=false,
 breaklinks=false,pdfborder={0 0 1},backref=false,colorlinks=true]
 {hyperref}
\usepackage{breakurl}

\makeatletter

\DeclareRobustCommand{\greektext}{%
  \fontencoding{LGR}\selectfont\def\encodingdefault{LGR}}
\DeclareRobustCommand{\textgreek}[1]{\leavevmode{\greektext #1}}
\DeclareFontEncoding{LGR}{}{}
\DeclareTextSymbol{\~}{LGR}{126}

\usepackage{braket}

\makeatother

\begin{document}
\global\long\def\tdel{q}

\title{Transition to chaos in random neuronal networks}

\author{Jonathan Kadmon}

\affiliation{Racah Institute of Physics and the Edmond and Lily Safra Center for
Brain Sciences, The Hebrew University of Jerusalem, 9190401 Israel}

\author{Haim Sompolinsky}

\affiliation{Racah Institute of Physics and the Edmond and Lily Safra Center for
Brain Sciences, The Hebrew University of Jerusalem, 9190401, Israel}

\affiliation{Center for Brain Science, Harvard University, Cambridge MA 02138,
USA}
\begin{abstract}
Firing patterns in the central nervous system often exhibit strong
temporal irregularity and considerable heterogeneity in time averaged
response properties. Previous studies suggested that these properties
are outcome of the intrinsic chaotic dynamics of the neural circuits.
Indeed, simplified rate-based neuronal networks with synaptic connections
drawn from Gaussian distribution and sigmoidal non-linearity are known
to exhibit chaotic dynamics when the synaptic gain (i.e., connection
variance) is sufficiently large. In the limit of infinitely large
network, there is a sharp transition from a fixed point to chaos,
as the synaptic gain reaches a critical value. Near onset, chaotic
fluctuations are slow, analogous to the ubiquitous, slow irregular
fluctuations observed in the firing rates of many cortical circuits.
However, the existence of a transition from fixed point to chaos in
neuronal circuit models with more realistic architectures and firing
dynamics has not been established.

In this work we investigate rate based dynamics of neuronal circuits
composed of several subpopulations with randomly diluted connections.
Nonzero connections are either positive-for excitatory neurons, or
negative for inhibitory ones, while single neuron output is strictly
positive with output rates rising as a power law above threshold;
in line with known constraints in many biological systems. Using Dynamic
Mean Field Theory, we find the phase diagram depicting the regimes
of stable fixed point, unstable dynamic and chaotic rate fluctuations.
We focus on the latter and characterize the properties of systems
near this transition. We show that dilute excitatory-inhibitory architectures
exhibit the same onset to chaos as the single population with Gaussian
connectivity. In these architectures, the large mean excitatory and
inhibitory inputs dynamically balance each other, amplifying the effect
of the residual fluctuations. Importantly, the existence of a transition
to chaos and its critical properties depend on the shape of the single-
neuron nonlinear input-output transfer function, near firing threshold.
In particular, for nonlinear transfer functions with sharp rise near
threshold, the transition to chaos disappears in the limit of a large
network; instead, the system exhibits chaotic fluctuations even for
small synaptic gain. 

Finally, we investigate transition to chaos in network models with
spiking dynamics. We show that when synaptic time constants are slow
relative to the mean inverse firing rates the network undergoes a
transition from fast spiking fluctuations with constant rates to a
state where the firing rates exhibit chaotic fluctuations, similar
to the transition predicted by rate based dynamics. Systems with finite
synaptic time constants and firing rates exhibit a smooth transition
from a regime dominated by stationary firing rates, to a regime of
slow rate fluctuations. This smooth crossover obeys scaling properties,
similar to crossover phenomena in statistical mechanics. The theoretical
results are supported by computer simulations of several neuronal
architectures and dynamics. Consequences for cortical circuit dynamics
are discussed. These results advance our understanding of the properties
of intrinsic dynamics in realistic neuronal networks and their functional
consequences.
\end{abstract}
\maketitle

\section{Introduction}

The firing patterns of circuits in the central nervous system often
exhibit a high level of temporal irregularity. The effect can be seen
by the inter-spike interval (ISI) distribution, which, except for
a short refractory period, is similar to that of a Poisson process
\cite{Abeles1991,BariKoch1994,Swift1993}. Intracellular recordings
\cite{Holt1996} indicate that this irregularity is due to fluctuations
in the synaptic input to the neurons, suggesting a dynamic origin,
and motivating the exploration of underlying neuronal circuit mechanisms.
A possibly related issue is the ubiquitous diversity of the time averaged
response properties of single neurons, e.g., their firing rates and
tuning modulations, within a local population \cite{Abeles1991,Kang:2004bw}. 

Several theoretical studies explored the emergence of temporal irregularity
and in particular chaotic dynamics in neuronal networks. These investigations
focused on two types of models: rate-based models with Gaussian connections,
and spiking dynamics of sparsely connected excitatory-inhibitory networks.
The first class uses a firing rate dynamics in which each unit is
characterized by a smooth function that maps the synaptic input into
an output firing rate. In its simplest version, the input-output transfer
function is $\tanh(x)$ where a zero value denotes some reference
activity level and $1$ and $-1$ denote saturated firing and quiescent
states, respectively. The architecture of the rate model was given
by a random connectivity matrix where each connection is drawn from
a Gaussian distribution, with zero mean and variance given by $g^{2}/N$,
$N$ being the size of the network. It was shown that the system exhibits
a transition from a stable zero fixed point state for low value of
$g$ to chaotic state for large $g$. Furthermore, for large $N$
this transition is sharp and occurs at $g_{c}=1$ \cite{Sompolinsky:1988ta}.
In these models, the emergence of chaos is gradual, as both the amplitude
of the fluctuations, their inverse time constant, and the Lyapunov
exponent vanish as $g\rightarrow1^{+}$. The chaotic state is asynchronous
in that the correlations between fluctuations of different neurons
are weak (and vanish as $N\rightarrow\infty$ ).

The second class is motivated by biological reality. To capture the
spiking dynamics, these models use either binary $\{0,1\}$ neurons,
or integrate-and-fire spiking neurons, and the connectivity is characterized
by randomly sparse connections, where the mean number of connections,
$K$, is much smaller than $N$. To capture the biological constraints
on the sign of the connections, the networks consist of excitatory
and inhibitory populations, where the nonzero output connections of
excitatory (inhibitory) neurons are positive (negative). The dynamics
is dominated by the competition between strong excitatory and inhibitory
connections, leading to a dynamic cancellation of the excitatory and
inhibitory inputs. The ensuing balanced state exhibits intrinsically
generated Poisson-like stochasticity as well as asynchrony \cite{vanVreeswijk:1996us,Vreeswijk:1998uz,Renart:2010hj,Brunel2000307}.
There is no parameter regime where the state can be characterized
as a stable fixed point, and chaos does not emerge gradually as a
function of synaptic strength. Instead, temporal irregularity is always
strong, and correlation times are short. The origin of the qualitative
difference in the behavior of the two types of models was never investigated
comprehensively. In particular, it was unclear whether the differences
originate from the different dynamics (a smooth rate dynamics vs.
spiking dynamics) or it is attributed to the different architectures:
Gaussian distributed synapses in a single, fully connected, and statistically
homogeneous population (with mixed excitation and inhibition) vs.
two population (excitatory and inhibitory) architecture with sparse
connectivity.

The question of the existence of a bifurcation to chaos is not only
interesting from dynamical systems perspective but may also have important
functional consequences. Several studies have highlighted the computational
utility of the nonlinear dynamics of random, recurrent networks near
the onset of chaos. For instance, a novel 'reservoir computing' model
has been proposed, which utilizes the rich intrinsic network dynamics
to learn to generate complex temporal trajectories \cite{Busing:2010hw,LukosEvicIus:2009bt,Sussillo2009544,Barak:2013bg,Legenstein:2007bg,Maass:2002kf,Jaeger:2004hj}.
Reservoir computing is most effective above but near the transition
to chaos, due to the emergence of slow dynamical fluctuations, since
many applications involve dynamics with time scales of seconds, much
larger than the microscopic time scales of a few milliseconds. Additionally,
it has been shown that decoding signals from these networks is particularly
robust above and near the transition to chaos \cite{Toyoizumi:2011vf,Bertschinger:2004cm}.
A recent study by Saxe \emph{et al} \cite{Saxe:2013tq} studies
the dynamics of learning in deep networks. They define a critical
point above which infinitely deep networks exhibit chaotic percolating
activity propagation, analogous to the chaotic state of recurrent
networks. Finally, recent advances in machine learning has generated
resurgence of interest in recurrent networks (primarily of speech
and language processing, e.g., \cite{Graves:dv,Mesnil:2015ip,pascanu2013construct}
and references therein). Understanding the dynamics of generic recurrent
networks will gain insight into these highly interesting computational
capacities. 

In this work, we study the existence and the properties of the transition
to chaotic dynamics in a broad range of models that span the two above
mentioned model classes. In section \ref{sec:Model} we introduce
a general architecture for random recurrent networks with multiple
sub-populations, obeying smooth rate based dynamics. We show the correspondence
between randomly diluted networks in the balanced state and networks
with Gaussian distributed connections with the same multiple population
architecture. Section \ref{sec:DMFT} introduces the mathematical
framework of the dynamic mean field theory (DMFT) used in analytical
investigation of the properties of the network state and extend the
theory of transition from fixed point to chaos, previously derived
for a single gaussian population to the more general architecture.
In section \ref{sec:Single-inhibitory-population} we apply the theory
to the simple case of a single inhibitory neuronal population and
a threshold linear\emph{ }synaptic transfer function. The example
of two population model is studied in section \ref{sec:Two-populations-1}.
Although the DMFT is more complex than the single population network,
we show that the two population network exhibits a transition from
fixed point to chaos, that is similar to that of the single population
case. The role of the single neuron nonlinear transfer function is
elucidated in \ref{sec:Criticality}. First, we show that for sufficiently
sharp function, chaos exists in large systems at all gain values.
Secondly, the shape of this function determines the critical behavior
of the relaxation times and largest Lyapunov exponent near the transition. 

Unlike the rate-based smooth dynamics, in models with spiking dynamics,
a fixed point state does not exist as long as some of the neurons
are firing. Nevertheless, it is interesting to explore the conditions
under-which the spiking network exhibits a transition from a state
with stationary inputs and firing rates, to a state where the underlying
inputs and rates fluctuate in time. In section \ref{sec:Spiking}
we use a Poisson spiking model, to show that in the case where the
synaptic integration time is much larger than the inverse of the neuronal
firing rates, the synaptic current exhibits a sharp transition from
a fixed point to chaotic dynamics, similar to that of a rate dynamics.
The behavior for large but finite synaptic integration time is analyzed
using scaling analysis, similar to that of a second order phase transition.
The implications of the results for the understanding of the dynamics
and computations in cortical circuits are discussed in section \ref{sec:Discussion}.

\section{Model\label{sec:Model}}

\subsection{Randomly diluted network with multiple populations}

\begin{figure}
\includegraphics[width=0.9\columnwidth]{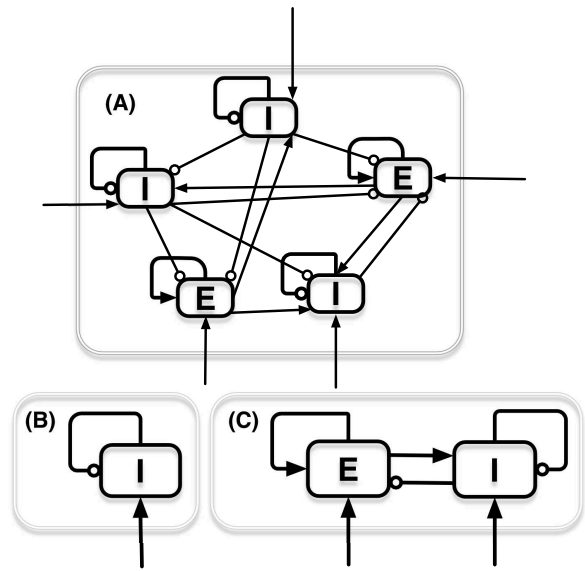}\protect\caption{\emph{Network schematics}. Arrows (circles) denote excitatory (inhibitory)
connections. \textbf{(A)} The general network architecture studied
here consists of multiple excitatory (E) and inhibitory (I) subpopulations
driven by external inputs. Individual connections are randomly diluted
or are Gaussian distributed. Subpopulations are distinguishable by
the different statistics of connectivity. \textbf{(B)} The simplest
model consists of a single population with random inhibitory recurrent
connections with an external excitatory input. \textbf{(C)} Two interconnected
randomly connected E and I populations. \label{fig:Schematics}}

\end{figure}

We consider a network of neurons composed of $P$ subpopulations which
are assumed for convenience to have equal size, $N$ (Fig. \ref{fig:Schematics}A).
The recurrent connections, $J_{kl}^{ij}$, denote the synaptic efficacy
between the presynaptic $j$th neuron of the $l$th population to
the post synaptic $i$th neuron of the $k$th population, where $k,l=1,...,P$,
and $i,j=1,...,N$. The connectivity is randomly diluted, so that
each connection $J_{kl}^{ij}$ is nonzero with probability $p,$ where 

\begin{equation}
p=K/N
\end{equation}
and zero otherwise. Thus, the mean number of inputs to each neuron
is $K$. Some of the populations are assumed to be inhibitory. For
an inhibitory population $l$, all $J_{kl}^{ij}$ are nonpositive.
Conversely, all nonzero outgoing connections of an excitatory populations
are positive. In addition, each neuron from the $k$-th population
receives a constant, uniform input, equal to $m_{0}W_{k}$, where
the parameter $m_{0}$ denotes the mean activity (firing rate) of
neurons in the input population, and $W_{k}$ are assumed positive. 

Throughout this work, we focus on networks with high degree of connectivity,
i.e., $N,K\gg1$. Although in most previous analytical work on the
dynamics of dilute neuronal networks it was assumed that the network
connectivity is sparse (i.e., $1\ll K\ll N$ ) \cite{Brunel2000307,Brunel:2000th,Vreeswijk:1998uz,Ostojic:2014kd},
here we will assume only that $1\ll K<N$ allowing for dense regime
as well (in \cite{Renart:2010hj,Helias:2014by} densely connected
network were studied, but the focus was on the spatial patterns and
correlations. The dynamical properties and the temporal autocorrelations
were not addressed).  In order for the dynamics to be affected by
the fluctuations in connectivity, we assume all nonzero connections
equal $J_{kl}/\sqrt{K}$, as discussed below. Similarly, the external
connections scale as as $W_{k}=\omega_{k}\sqrt{K}$. 

\subsection{Rate dynamics}

We first study a firing rate model for the dynamics of neuronal activity.
The state of each neuron, say, with indices $(i,k)$, at time $t$,
is given by its total synaptic \emph{current}, $h_{k}^{i}(t)$, that
obeys a first order nonlinear dynamics, 

\begin{eqnarray}
\frac{d}{dt}h_{k}^{i}(t) & = & -h_{k}^{i}(t)+\sum_{l=1}^{P}\sum_{j=1}^{N}J_{kl}^{ij}\phi\left(h_{l}^{j}(t)\right)+W_{k}m_{0}.\label{eq:Full_model}
\end{eqnarray}
The first term is a decay term, the second is the \emph{synaptic}
\emph{input }from the network and the last is the synaptic input from
the external source. For convenience we set all synaptic time constants
to unity. The nonlinear transfer function $\phi(h)$ denotes the firing
rate of a neuron with an input synaptic current, $h$, and is analogous
to the neuronal input-current to firing-rate transfer function, known
as the \emph{f-I curve}.

\subsection{Effective gaussian connectivity}

Past work on random highly connected systems has shown that under
fairly general conditions, in the limit of large $K$, the system's
behavior depends only on the first two moments of the connectivity
matrix \cite{Hertz_SG,mezard1987spin}. Hence the connectivity matrix
$J_{kl}^{ij}$ can be replaced by a random matrix of Gaussian distributed
connections where the mean and variance are matched to that of the
dilute network. We thus consider a general dynamics of a multiple
population network with fully connected connectivity matrix with Gaussian
distributed connections, $J_{kl}^{ij}=\frac{\bar{g}_{kl}}{N}+\mathcal{J}_{kl}^{ij}$, 

\begin{multline}
\frac{d}{dt}h_{k}^{i}(t)=-h_{k}^{i}(t)+\sum_{l=1}^{P}\sum_{j=1}^{N}\mathcal{J}_{kl}^{ij}\phi\left(h_{l}^{j}(t)\right)\\
+\sum_{l=1}^{P}\bar{g}_{kl}m_{l}(t)+h_{k}^{0}\label{eq:Full_model_gauss}
\end{multline}
where

\begin{equation}
m_{k}(t)=\frac{1}{N}\sum_{i=1}^{N}\phi\left(h_{k}^{i}(t)\right)\label{eq:popM}
\end{equation}
are the mean population activities. The coefficients $\mathcal{J}_{kl}^{ij}$
are quenched Gaussian variables with zero mean and variance

\begin{equation}
\langle(\mathcal{J}_{kl}^{ij})^{2}\rangle=\frac{g_{kl}^{2}}{N}.\label{eq:J_statistics}
\end{equation}
We refer to $g$ as the gain of the synaptic input. The contributions
of the mean connections, $\frac{\bar{g}_{kl}}{N}$, is represented
by the third term in the RHS of Eq (\ref{eq:Full_model_gauss}) .

In the dilute network, the mean connection between two populations
equals $pJ_{kl}/\sqrt{K}=\sqrt{K}J_{kl}/N$. The variances of these
connections are $p(1-p)J_{kl}^{2}/K=(1-K/N)J_{kl}^{2}/N$. Hence the
corresponding parameters of the equivalent Gaussian network are, 

\begin{equation}
g_{kl}^{2}=(1-K/N)J_{kl}^{2},\;\bar{g}_{kl}=\sqrt{K}J_{kl},\;h_{k}^{0}=\sqrt{K}\omega_{k}m_{0}.\label{eq:sqrtK}
\end{equation}

Note that in the dilute networks, $g_{kl}$ and $\bar{g}_{kl}$ are
related through

\begin{equation}
\bar{g}_{kl}=g_{kl}\sqrt{\frac{K}{(1-K/N)}}.\label{eq:gg0}
\end{equation}

In the theory below, we will analyze the dynamics of the Gaussian
network, defined by Eq (\ref{eq:Full_model_gauss}), in its generality;
namely, we will consider the parameters describing the mean and the
variance of the connections, $\bar{g}_{kl}$ and $g_{kl}$, as independent
parameters. In addition, we will not restrict ourselves to the case
that the mean connections $\bar{g}_{kl}$ and $h_{k}^{0}$ are large.
The relation, (\ref{eq:gg0}), as well as the scaling of $\bar{g}_{kl}$
and $h_{k}^{0}$ with $O(\sqrt{K}$) will be adopted when applying
the theory to the balanced dilute networks. The numerical simulations
will use both Gaussian and random dilution and will show their equivalence.

\subsection{The balanced regime}

In the diluted network model, the net input from the $l$-th population
to each of the $k$-th neurons, is proportional to $\bar{g}_{kl}$,
and $h_{k}^{0}$, hence scales as $\sqrt{K}$. On the other hand,
the fluctuations scale as $g_{kl}$, hence are of order $1$. Thus,
it would seem that whenever the degree of connectivity, $K$, is high,
the fluctuations induced by the random dilution will have negligible
effect on the dynamics, reducing the system to that with a uniform
connections and anomalously strong recurrent and external inputs.
In fact, the system avoids this saturated state, by dynamically canceling
the mean excitatory contributions against the inhibitory ones, resulting
in the net mean inputs which are of order $1$, and of the same order
as the fluctuations. Balanced states in networks with strong excitatory
and inhibitory connections were previously studied (see \cite{vanVreeswijk:2005uo,Vreeswijk:1998uz}
) in the context of binary or spiking networks. As we will show below,
networks of units with smooth rate based dynamics and transfer functions
with finite gains also settle in this balanced state.

\section{Fixed points and their stability\label{sec:DMFT}}

\subsection{Mean field equations for the fixed points}

To fully characterize the state of the system we separate the population
averaged quantities from the fluctuating ones, writing, 

\begin{equation}
h_{k}^{i}(t)=u_{k}+\delta h_{k}^{i}(t)
\end{equation}

where, 

\begin{equation}
u_{k}=\sum_{l}\bar{g}_{kl}m_{l}+h_{k}^{0}\label{eq:uk}
\end{equation}

The fluctuating inputs obey 

\begin{equation}
\frac{d}{dt}\delta h_{k}^{i}(t)=-\delta h_{k}^{i}(t)+\eta_{k}^{i}(t).\label{eq:mft}
\end{equation}

where the spatiotemporal fluctuations in the currents, $\delta h_{k}^{i}(t)$,
are low pass temporal filters of the fluctuating synaptic inputs,

\begin{equation}
\eta_{k}^{i}(t)=\sum_{l}\sum_{j}\mathcal{J}_{kl}^{ij}\phi\left(h_{l}^{j}(t)\right).\label{eq:mft_noise}
\end{equation}
In the limit of a large number of inputs per neuron, these quantities
obey Gaussian statistics with zero mean, zero spatial correlation
but with temporal correlations which need to be evaluated self-consistently
as explained below. 

First, we study the fixed point solution of the network dynamics,
corresponding to a time independent state, where $\delta h_{k}^{i}(t)=\delta h_{k}^{i}=\delta\eta_{k}^{i}$
are Gaussian distributed with zero mean and variance $\Delta_{k}\equiv\langle(\delta h_{k}^{i})^{2}\rangle$.
This quantity is calculated self-consistently, as 
\begin{equation}
\Delta_{k}=\langle(\eta_{k}^{i})^{2}\rangle=\sum_{l}g_{kl}^{2}C_{l},\label{eq:FP_delta0}
\end{equation}

where $C_{l}=\langle\phi^{2}(h)\rangle$ are the average autocorrelations
of the neuronal activities. Finally using $h_{k}^{i}=u_{k}+\delta\eta_{k}^{i},$we
obtain the self consistent equation for $\Delta_{k}$

\begin{equation}
C_{k}=\left\langle \phi^{2}\left(\sqrt{\Delta_{k}}z+u_{k}\right)\right\rangle .
\end{equation}

The constants $u_{k}$ are determined self-consistently, via Eq (\ref{eq:uk}),
and

\begin{equation}
m_{k}=\left\langle \phi\left(\sqrt{\Delta_{k}}z+u_{k}\right)\right\rangle ,\label{eq:mfM}
\end{equation}

Note that in Mean Field Theory all averages denote integration over
the Gaussian variables ($z$ in the above equations) which have zero
means and unit variance.

\subsection{The balance equations}

In balanced architectures both $\bar{g}_{kl}$ and $h_{k}^{0}$ are
of order $\sqrt{K}$ (see Eq (\ref{eq:sqrtK})). In this case, the
self consistent equations for $m_{k}$ assume a simple form. This
is because for the system to settle into an unsaturated state, $u_{k}$
must be of order $1$; hence Eq (\ref{eq:uk}) yield (to leading order
in $\sqrt{K}$)

\begin{equation}
\sum_{l}\bar{g}_{kl}m_{l}+h_{k}^{0}=0,\;\forall k.\label{eq:balance}
\end{equation}

Since the mean rates are non-negative, Eq (\ref{eq:balance}) can
be obeyed only for a range of $\bar{g}_{kl}$ and $h_{k}^{0}$ values.
Stability of the balanced state (see \cite{Vreeswijk:1998uz}) further
restricts the parameter regimes. We refer to these restrictions on
the parameters as the \emph{balance conditions}. Substituting the
solution to the balance equations into Eq (\ref{eq:mfM}) yields
equations for the residual, order $1$ mean inputs, $u_{k}$.

\subsection{Stability of fixed points}

\paragraph{Stability of the population average activities}

The stability equations for the population averaged degrees of freedom
are determined by considering the response of the system to perturbations
in the external fields, $\delta h_{k}^{0}$ , which are uniform within
the populations. It is convenient to define the uniform linear response
by 

\begin{equation}
\chi_{kl}(t)\equiv\frac{\partial m_{k}(t)}{\partial h_{l}(0)}=\frac{1}{N}\sum_{i}^{N}\partial\phi_{k}^{i}(t)/\partial h_{l}^{0}(0).
\end{equation}

Note that due to the spatial summation, this quantity is essentially
averaged over $\mathcal{J}$. Interestingly, the response of the population
averaged activity is coupled to the response of the population variances,
defined as:

\begin{equation}
\chi_{kl}^{\Delta}(t)=\frac{\partial}{\partial h_{l}^{0}(0)}\Delta_{k}(t)=\frac{\partial}{\partial h_{l}^{0}(0)}\langle\delta h_{k}^{i}(t)\delta h_{k}^{i}(t)\rangle.
\end{equation}
In the Appendix \ref{sec:App:UniformStability} we derive the following
coupled equations for the two sets of susceptibilities in the temporal
Fourier domain,

\begin{equation}
\begin{bmatrix}(\mathbf{I}-\mathbf{A}\bar{\mathbf{g}}+i\omega) & \mathbf{-B}\\
-\mathbf{E}\bar{\mathbf{g}} & (\mathbf{I}-\mathbf{D}+i\omega)
\end{bmatrix}\begin{bmatrix}\chi(\omega)\\
\chi^{\Delta}(\omega)
\end{bmatrix}=\begin{bmatrix}\mathbf{A}\\
\mathbf{E}
\end{bmatrix}\label{eq:uniformstabeq}
\end{equation}

Here, $\chi$ and $\chi^{\Delta}$are both $PxP$ ($P$ being the
number of populations). The $PxP$ dimensional matrices appearing
in (\ref{eq:uniformstabeq}) are defined as, 

\begin{alignat*}{1}
A_{kl} & =g_{kl}^{2}\langle\phi_{l}\phi{}_{l}'\rangle+\langle\phi{}_{l}'\rangle\\
B_{kl} & =\frac{1}{2}g_{kl}^{2}\langle\phi_{k}''\rangle[\langle(\phi{}_{l}')^{2}\rangle+\langle\phi_{l}\phi{}_{l}''\rangle]\\
D_{kl} & =g_{kl}^{2}\left(\langle(\phi{}_{l}')^{2}\rangle+\langle\phi_{l}\phi{}_{l}''\rangle\right)\\
E_{kl} & =2g_{kl}^{2}\langle\phi_{l}\phi_{l}'\rangle
\end{alignat*}

Thus, the fixed point is stable against population average perturbations
provided that all the eigenvalues of the matrix 
\begin{equation}
\begin{bmatrix}\mathbf{I}-\bar{\mathbf{g}}\mathbf{A} & \mathbf{-B}\\
-\mathbf{\bar{g}E} & \mathbf{I}-\mathbf{D}
\end{bmatrix}\label{eq:uniformStabDet}
\end{equation}

have negative real part.

\paragraph{Stability against local perturbations}

We now study the stability of the fixed point against small perturbations
in the form of infinitesimal local fields $h_{k}^{0i}$ . It is convenient
to define the local susceptibility matrix 
\begin{equation}
\chi_{kl}^{ij}(t)=\partial h_{k}^{i}(t)/\partial h_{l}^{j0}(0)
\end{equation}

The average of the off-diagonal elements of this matrix is zero, and
their variance is $\mathcal{O}(1/N)$, hence we focus on the mean
square susceptibility matrix, $\mathbf{G}$, defined in the Fourier
domain by

\begin{equation}
G_{kl}(\omega_{1,},\omega_{2})=N^{-1}\sum_{ij}\langle\chi_{kl}^{ij}(\omega_{1})\chi_{kl}^{ij}(\omega_{2})\rangle\label{eq:localG}
\end{equation}

In Appendix \ref{sec:Appendix_Stability} we show that the matrix
$\mathbf{G}$ is

\begin{equation}
\mathbf{G}(\omega_{1,},\omega_{2})=[(1+i\omega_{1})(1+i\omega_{2})I-\mathbf{M}]^{-1}\label{eq:stabFP}
\end{equation}
where 
\begin{equation}
M_{kl}=g_{kl}^{2}\left\langle (\phi'){}^{2}\right\rangle .\label{eq:stability_matrix_M}
\end{equation}
Thus, the fixed points are stable provided the real parts of all eigenvalues
of the $P$ dim matrix $\mathbf{M}$ are all less than $1$. In general,
$\mathbf{M}$ is not symmetric; however, the largest (in absolute
value) of the eigenvalues is real. We will call $\mathbf{M}$ the
\emph{stability matrix}.

In conclusion, we have derived two stability conditions: one related
to population average perturbations, Eq (\ref{eq:uniformStabDet}),
and a second related to local perturbations, associated with (\ref{eq:stabFP}).
Note that the mean interactions $\bar{g}$ do not appear in the latter
condition. This is because the contribution to the off-diagonal elements
of (\ref{eq:localG}) from the uniform susceptibility is only of the
order $1/N$ . For the fixed point to be stable, both conditions must
hold. However the instability associated with each condition has a
different implication. The population average instability signals
either a runaway (as we will show in a concrete example in Section
\ref{sec:Single-inhibitory-population}) or a transition to another
stable fixed point, a stable limit cycle, or some other coherent spatio-temporal
states. On the other hand, as we will show below, the instability
in (\ref{eq:stabFP}) signals a transition an asynchronous chaotic
state. 

Which of the instabilities occurs first when one varies one of the
parameters depends on the specific architecture and parameter sets.
Specific examples will be shown below.

\section{Chaotic state: Dynamic Mean Field Theory}

The chaotic state is an asynchronous state with stationary statistics,
governed by the local-field autocorrelation functions

\begin{equation}
\Delta_{k}(\tau)\equiv\langle\delta h_{k}^{i}(t)\delta h_{k}^{i}(t+\tau)\rangle,\label{eq:Autocorrelations}
\end{equation}

where the angular brackets denote both average over neurons in the
population and over time $t$. The self-consistent equations for these
functions, derived from the Dynamic Mean Field Theory (DMFT) are (see
Appendix \ref{sec:app:DMFT-derivation} for detailed derivation):

\begin{equation}
\left(1-\frac{\partial^{2}}{\partial\tau^{2}}\right)\Delta_{k}(\tau)=\sum_{l}g_{kl}^{2}C_{l}(\tau),\label{eq:dMFT_delta}
\end{equation}

where $C_{k}(\tau)=\langle\phi(h_{k}^{i}(t))\phi(h_{k}^{i}(t+\tau))\rangle$
are the firing rate autocorrelation functions, which depends on $\Delta_{k}(\tau)$
through

\begin{equation}
C_{k}(\tau)=\left\langle \left\langle \phi\left(\sqrt{\Delta_{k}^{0}-\Delta_{k}(\tau)}y+\sqrt{\Delta_{k}(\tau)}z+u_{k}\right)\right\rangle _{y}^{2}\right\rangle _{z}.\label{eq:dMFT_C}
\end{equation}
Here, both $y$ and $z$ are Gaussian distributed random variables
with zero means and unit variances. The boundary conditions for the
solution are: $\partial\Delta_{k}(0)/\partial\tau=0$; $\partial\Delta_{k}^{2}(\infty)/\partial\tau^{2}=0$;
and $\Delta_{k}(0)=\Delta_{k}^{0}$ .The first condition stems from
the general fact that $\Delta(\tau)$ is a symmetric, continuous function.
The second one comes from the requirement that $\Delta(\tau)$ converges
to a finite value as $\tau\rightarrow\infty$ for a chaotic solution.
Solution of (\ref{eq:dMFT_C}) subject to these boundary conditions
yields a unique solution for $\Delta_{k}(\tau)$ that converges to
a fixed value, $\Delta_{k}(\infty)$ at long time. Except for networks
with $h\rightarrow-h$ symmetry, the quantity $\Delta_{k}(\infty)$
is in general not zero, and represents the variance of the Gaussian
distribution of the \emph{time averaged} synaptic currents. The fluctuations
in these inputs determine the fluctuations in the time averaged firing
rates of individual neurons. 

Details of the derivation of the DMFT are given in appendix \ref{sec:app:DMFT-derivation}.
Alternative formalisms for deriving dynamic mean field equations for
such architectures are Faugeras and coworkers using Mckean-Vaslov
Fokker Plank formalisms \cite{faugeras2009constructive}, or for stochastic
networks via path integrals (e.g., \cite{Sompolinsky:1982jg,buice2010systematic}).

\paragraph{Lyapunov exponent}

In the chaotic state, we expect the squared susceptibility to show
an exponential divergence with time with a rate constant given by
the largest Lyapunov exponent (LE). LE is defined as 

\begin{equation}
\lambda_{L}=\lim_{\tau\to\infty}\frac{1}{2\tau}\ln\sum_{kl}\left(G_{kl}(\tau)\right),\label{eq:LE_theory}
\end{equation}

where 

\begin{equation}
G_{kl}(\tau)=\lim_{t\to\infty}\frac{1}{N}\sum_{i,j}^{N}\left(\frac{\partial h_{k}^{i}(t+\tau)}{\partial h_{l}^{0j}(t)}\right)^{2}.\label{eq:Chi_square}
\end{equation}

Extending previous calculations \cite{Sompolinsky:1988ta}, we show
in appendix \ref{sec:Appendix_Stability} that the long time behavior
of $\mathbf{G}(\tau)$ is determined by the ground state of a quantum
mechanical problem with the Hamiltonian operator defined as 
\begin{equation}
\mathcal{H}=-\frac{\partial^{2}}{\partial\tau^{2}}\mathbf{I}+\mathbf{I}-\mathbf{M}(\tau),\label{eq:Hamiltonian}
\end{equation}

where 

\begin{equation}
M_{kl}(\tau)=g_{kl}^{2}\left\langle \left\langle \phi'\left(\sqrt{\Delta_{l}^{0}-\Delta_{l}(\tau)}y+\sqrt{\Delta(\tau)}z+u_{l}\right)\right\rangle _{y}^{2}\right\rangle _{z}.\label{eq:M_kl-def}
\end{equation}

Finally, the LE is equal to
\begin{equation}
\lambda_{L}=-1+\sqrt{1-\epsilon_{0}},\label{eq:Lambda_max}
\end{equation}

where $\epsilon_{0}$ is the ground state energy of $\mathcal{H}$.
Note that, in general, $\mathcal{H}$ is not Hermitian. However, complex
values of the Lyapunov exponents lead to unphysical oscillations of
the squared susceptibility and so the ground state energy is expected
to be real and negative. Direct rigorous proof is still missing. In
the regime of a stable fixed point, $\mathbf{M}$ is time independent
and $\epsilon_{0}>0$, recovering the same stability criterion as
above. A transition from a stable fixed point to chaotic dynamics
occurs as $\epsilon_{0}$ vanishes.

\section{Single inhibitory population with threshold- power law transfer function\label{sec:Single-inhibitory-population}}

Solving the DMFT for general networks requires extensive numerics.
Many of the salient features are captured by the simplest case of
a single inhibitory population (Fig. \ref{fig:Schematics}B) driven
by a constant external excitatory input. The systems's dynamics,

\begin{equation}
\frac{d}{dt}h^{i}(t)=-h^{i}(t)+\sum_{j=1}^{N}\mathcal{J}^{ij}\phi\left(h^{j}(t)\right)+\bar{g}m(t)+h^{0},\label{eq:one_pop_circuit_eq}
\end{equation}
is characterized by the recurrent inhibitory mean gain parameter $\bar{g}<0$,
the gain parameter of the synaptic fluctuations, $g^{2}=N\langle(\mathcal{J}^{ij})^{2}\rangle$,
and the excitatory external input, $h^{0}>0$. The system is further
simplified by assuming a transfer function of a threshold- power law
form \cite{Priebe:2008kn,Hansel:2002vz}, 
\begin{equation}
\phi(x)=[x]_{+}^{\nu},\;\nu>0,\label{eq:phi_nu}
\end{equation}
where $[x]_{+}=\max(x,0)$. For reasons explained below, we will see
that the this monomial form is general enough when one studies the
properties of the chaotic instability in networks with continuous
threshold transfer functions.

\subsection{Fixed point and its stability}

In this model, the mean field equations for the fixed point are particularly
simple. First, the mean input and its variance are given respectively,
by

\begin{equation}
u=\bar{g}m+h^{0},\label{eq:u}
\end{equation}

where $m$ is the spatially averaged activity, $\left\langle \phi_{i}\right\rangle $,
and

\begin{equation}
\Delta=\left\langle \delta h_{i}^{2}\right\rangle =g^{2}C,\label{eq:SP_Variance}
\end{equation}
where $\delta h_{i}=h_{i}-u$ and $C$ is the mean square activity,
$\left\langle \phi_{i}^{2}\right\rangle $.

The mean field equations for $m$ and $C$ are given by

\begin{equation}
m=\Delta^{\nu/2}\langle[z+x]_{+}^{\nu}\rangle,\label{eq:m}
\end{equation}
and
\begin{equation}
C=\Delta^{\nu}\langle[z+x]_{+}^{2\nu}\rangle.\label{eq:Cpowerlaw}
\end{equation}
Here 
\begin{equation}
x=u/\sqrt{\Delta},\label{eq:x_def}
\end{equation}
representing the mean input in units of the standard deviation, and
we have used the homogeneity property of the transfer function. 

Substituting (\ref{eq:Cpowerlaw}) in Eq (\ref{eq:SP_Variance}),
the parameter $x$ can be determined from 
\begin{equation}
1=\tilde{g}^{2}\left\langle [z+x]_{+}^{2\nu}\right\rangle ,\label{eq:SP_FP_sol}
\end{equation}
where $\tilde{g}=g\Delta^{(\nu-1)/2}$. The values of $m$ and $\Delta$
are evaluated using Eqs (\ref{eq:u}), (\ref{eq:x_def}), and (\ref{eq:m}). 

As one increases $g$ , the fixed point becomes unstable. Whether
the first instability is the chaotic one, (Eq (\ref{eq:stabFP}))
or the uniform one (Eq (\ref{eq:uniformstabeq})), will generally
depend on the parameter set ($\nu,$ $\bar{g}$ and $h^{0}$).

\paragraph{Chaotic Stability}

A transition to a chaotic state occurs at the fixed point state when 

\begin{equation}
1=\tilde{g}^{2}\nu^{2}\left\langle \left[z+x\right]_{+}^{2(\nu-1)}\right\rangle .\label{eq:SP_FP_stab}
\end{equation}
For a threshold-linear transfer function ($\nu=1$), $\tilde{g}=g$
is independent of $\Delta$ and critical values $g_{c}$ and $x_{c}$,
at which a bifurcation from a fixed point to chaotic dynamics occurs
is determined by Eq (\ref{eq:SP_FP_sol}) and (\ref{eq:SP_FP_stab})
and are independent of $h^{0}$. In this case $C'(x)=H(-x)$ with
$H(x)=\frac{1}{\sqrt{2\pi}}\int_{x}^{\infty}dze^{-\frac{z^{2}}{2}}$,
and transition occurs when $g_{c}=\sqrt{2},$and $x_{c}=0$. For non
linear transfer functions, the critical values $x_{c}$and $g_{c}$
depend on the values of $h^{0}$and $\bar{g}$.

\paragraph{Uniform Stability}

The stability condition for a network of single population against
uniform perturbations is given by (See Appendix \ref{sec:App:UniformStability}
where derive the uniform stability condition in Eq (\ref{eq:uniformstabeq})
for a single population for a single population):
\begin{equation}
g^{2}\left(\langle\phi'^{2}\rangle+\langle\phi\phi''\rangle+\frac{\langle\phi''\rangle\langle\phi\phi'\rangle\bar{g}}{(1-\langle\phi'\rangle\bar{g})}\right)<1.\label{eq:SingPop_Uniform_Stab}
\end{equation}
For $\nu=1$, $\langle\phi\phi''\rangle$ vanishes and the third term
in the LHS of (\ref{eq:SingPop_Uniform_Stab}) is negative for $\bar{g}<0$,
hence a chaotic instability (Eq \ref{eq:SP_FP_stab}) always occurs
first (i.e., for lower values of $g$). For $\nu\neq1$, the occurrence
of uniform stability of the fixed point will depend on the parameters
$\bar{g}$ and $h^{0}$.

\subsection{Chaotic state}

In the threshold-linear model, it is useful to define the normalized
autocorrelation

\begin{equation}
\tdel(\tau)\equiv1-\Delta(\tau)/\Delta(0),\label{eq:SP_q}
\end{equation}

as well as 

\begin{equation}
\tdel(\infty)=1-\Delta(\infty)/\Delta(0),\label{eq:q_inf}
\end{equation}

which measures the normalized variance of the dynamics, namely, the
(spatially averaged) temporal variance of the local fields normalized
by the squared amplitude. The autocorrelation function obeys (\ref{eq:dMFT_delta})
\begin{equation}
\left(1-\frac{\partial^{2}}{\partial\tau^{2}}\right)\Delta(\tau)=g^{2}C(\tau).
\end{equation}
 Normalizing by $\Delta_{0},$ and using the homogeneity of the transfer
function as above, we can define a Newtonian equation of motion on
the normalized autocorrelation (\ref{eq:SP_q}) which reads

\begin{equation}
\frac{\partial^{2}}{\partial\tau^{2}}\tdel(\tau)=-\frac{\partial V}{\partial q}.\label{eq:one_pop_EOM}
\end{equation}

The potential $V$ is given by

\begin{multline}
V\left(\tdel\right)\equiv-\frac{1}{2}\left(1-\tdel\right)^{2}+\\
\tilde{g}^{2}\intop_{-\infty}^{\infty}Dz\left[\intop_{-\infty}^{\infty}Dy\Phi\left(\sqrt{\tdel}y+\sqrt{(1-\tdel)}z+x\right)\right]^{2},\label{eq:one_pop_potential}
\end{multline}

where $\int Dx=\int dx\exp(x^{2}/2)/\sqrt{2\pi}$, and $\Phi(h)=\int_{0}^{h}dy\phi(y)=[h]_{+}^{\nu+1}/(\nu+1)$
\cite{Sompolinsky:1988ta}. In the above equation we have used the
equality $\Delta_{0}=\Delta(0)$ and $\tilde{g}\equiv g\Delta_{0}^{(\nu-1)/2}$.
The initial conditions for (\ref{eq:one_pop_EOM}) are $\tdel(0)=0$,
$\partial\tdel(0)/\partial\tau=0$, and $\partial^{2}\tdel(\infty)/\partial\tau^{2}=0$.
The second condition stems from the general fact that $\Delta(\tau)$
is a symmetric, continuous function. The last one comes from the requirement
that $\Delta(\tau)$ converges to a finite value as $\tau\rightarrow\infty.$
These three conditions yield a unique solution and a unique value
for the normalized mean input $x$. 

Due to the existence of a potential, $x$ can be obtained without
explicitly solving for $\tdel(\tau)$ as follows. The above boundary
conditions imply that the initial energy equals the potential energy,
$V(0)$, while the final energy equals the final potential energy,
$V(\tdel(\infty))$. Thus, conservation of total energy yields $V(0)=V(\tdel(\infty))$.
Finally, for $\tdel(\infty)$ to be an equilibrium point, the force
must vanish, hence $\partial V(\tdel(\infty))/\partial\tau=0$. These
two equations determine both $\tdel(\infty)$ and $x$. Once $x$
is known, (\ref{eq:one_pop_EOM}) is integrated numerically to yield
$\tdel(\tau)$. Finally, $\Delta(\tau)$ is evaluated by solving the
mean field equation (\ref{eq:m}) with $\Delta=\Delta_{0}$.

\paragraph{Stability of the chaotic solution}

The existence of a bounded chaotic phase, in which the mean activity
of the network does not diverge and the trajectories of the local
fields remain bounded requires stability of the uniform mode. Unfortunately
a theory of uniform stability in a time dependent dynamical state
is lacking. However, in the linear case, one can find simple arguments
for the existence of a chaotic solution. In this case, the normalized
mean input $x<0$ in the chaotic phase is independent of $h^{0}$,
and one finds that only when 
\begin{equation}
|\bar{g}|>\frac{|x|}{\left\langle [z+x]_{+}\right\rangle },\label{eq:one_pop_critical_g0}
\end{equation}
does a solution for the mean field equation exists; smaller values
of $|\bar{g}|$, entails dynamical instability. 

For non linear transfer functions, the above argument does not hold.
For $\nu=2$ for example, a solution for the MF equation exist for
any value of $h^{0}$and $\bar{g}$.  However, numerical simulations
show that an instability in the chaotic phase exist, as can be seen
in Fig. \ref{fig:Phase-diagram}.

We note that this instability results from the unboundedness of $\phi$.
For $\phi$ with a saturation level, the dynamics will always be bounded
but, a crossover is expected from fluctuations spanning the linear
dynamic range of the neurons when the net inhibition is large, to
'epileptic' fluctuations in which neurons fluctuate between their
saturated levels, for weak inhibition.

\paragraph{Existence of a fixed point solution}

An interesting result that is implied by Eq (\ref{eq:SP_FP_stab})
is that a stable FP exist only when 
\begin{equation}
\nu>1/2.
\end{equation}

In contrast, when $\nu\leq1/2,$ the RHS of the FP stability condition,
(\ref{eq:SP_FP_stab}), diverges indicating that no stable fixed point
solution exists for finite $g$, and depending on $g$, the system
is either in a stable chaotic state or diverges. This prediction is
confirmed by the numerical simulations, Fig. \ref{fig:Phase-diagram}A,
in which the normalized variance of the fields $\tdel(\infty)$ is
plotted as a function of $\nu$. For $\nu$ values of $0.5$ or smaller
the system is in a chaotic state even for small values of $g$. 

The instability of the fixed point for small $\nu$ is due to the
presence of positive local fields that are arbitrarily close to zero.
Thus, for a system of finite size, where the positive local fields
are always of a nonzero minimum value, we expect that a fixed point
will be stable at sufficiently small values of $g$. In the following
sections, we focus on networks with threshold- linear ($\nu=1$) and
quadratic ($\nu=2)$ transfer function, which exhibit a fixed point,
a chaotic regime and a transition therefore.

Finally we note that the same arguments hold for the multiple population
case as well.

\subsection{Phase diagram}

\begin{figure}
\includegraphics[width=1\columnwidth]{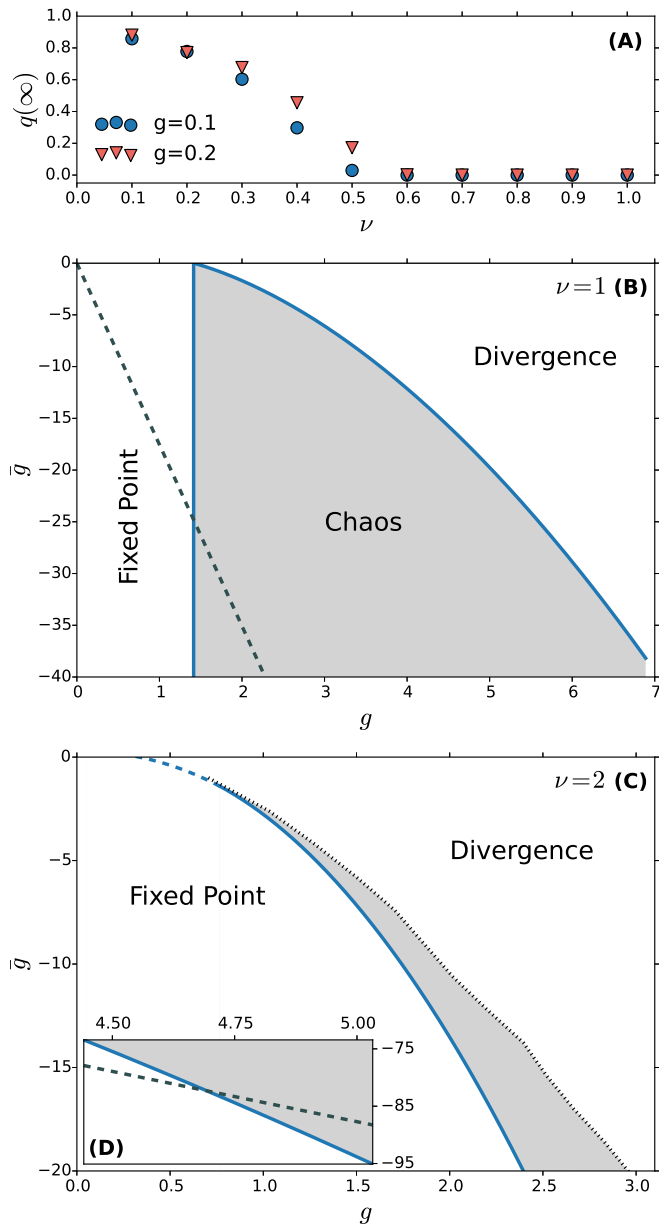}

\protect\caption{\emph{Phase diagrams for a single inhibitory population. }\textbf{\emph{(A)}}\emph{
No stable fixed point for sharp transfer functions.} Simulation results
for the normalized variance $\protect\tdel(\infty)$ for networks
with $g=0.2$ and $g=0.1$ and different values of the power-law,
$\nu$, characterizing the rise of the transfer function, Eq (\ref{eq:phi_nu}).
Below $\nu=0.5$ the variance is non-zero, implying there are temporal
fluctuations even for low values of $g$. Simulation preformed on
an inhibitory network ($N=5000$) with randomly diluted connections
and mean activity of $m=0.1$. \textbf{(B)} Phase space diagram for
a threshold-linear ($\nu=1$) network of an inhibitory neurons with
mean connectivity $\bar{g}<0$ and variance $g$ . Dashed line shows
an example for a transition in a diluted network (with $K=650$ )
where because $\bar{g}=-\sqrt{K}g$ (see Eq (\ref{eq:SingPop_BalanceExistance}))
the network lies on a line in the phase diagram. \textbf{(C)} Phase
diagram for a threshold-quadratic ($\nu=2)$ network with an excitatory
external field $h^{0}=1$. Dashed line show uniform instability of
the fixed point and solid line shows the transition to the chaotic
phase (shaded area). Dotted (black) line shows the instability of
the chaotic phase found by simulations (Gaussian connectivity, $N=6000$,
$h^{0}=1$). Inset \textbf{(D)} same as (C) for larger values of $\left|\bar{g}\right|$.
Dashed line shows existence line for a diluted network as in (B).\label{fig:Phase-diagram} }
\end{figure}

In Figure \ref{fig:Phase-diagram}B we show the phase diagram for
the threshold-linear ($\nu=1$), depicting the regimes of stable fixed
point, chaos, and unstable dynamics in the parameter space of $g$
and $\bar{g}$. For values of $g<\sqrt{2}$ the network settle into
a fixed point. For larger gain values, if the inhibition is strong
enough, i.e. the condition in (\ref{eq:one_pop_critical_g0}) holds,
then the dynamics is chaotic and bounded to a finite regime in the
state phase. For lower values of the uniform inhibition, the mean
activity diverges. Note that due to the semi-linearity of the transfer
function, the phase diagram is independent of the magnitude of $h_{0}$
as long as it is positive. In a diluted network, the phase-plane is
reduced to a single line 
\begin{equation}
\bar{g}=-g\sqrt{\frac{K}{1-K/N}}.\label{eq:SingPop_BalanceExistance}
\end{equation}
For a sparse network where $K\ll N$, $\bar{g}=-\sqrt{K}$ (dashed
line in \ref{fig:Phase-diagram}E). Thus, for large values of $K$,
as in the balanced network case, the chaotic state at $g>\sqrt{2}$
is always stable. 

In \ref{fig:Phase-diagram}C The phase diagram for the semi-quadratic
($\nu=2$) is presented for $h^{0}=1$. For low values of $\left|\bar{g}\right|$,
uniform instability occurs at lower $g$ than chaotic instability,
and no stable chaotic phase exists. For larger values of $\left|\bar{g}\right|$,
a critical transition between a fixed point and chaotic dynamics exist.
For larger values of gain $g$, the dynamics always diverges; this
instability is shown by numerical simulation in the phase diagram.
For a diluted network, where $\left|\bar{g}\right|=\mathcal{O}(\sqrt{K})$,
there is always a chaotic transition, as shown in the inset of Fig
\ref{fig:Phase-diagram}C.

\subsection{Analytical evaluation of the Lyapunov exponent}

\begin{figure}[h]
\includegraphics[width=1\columnwidth]{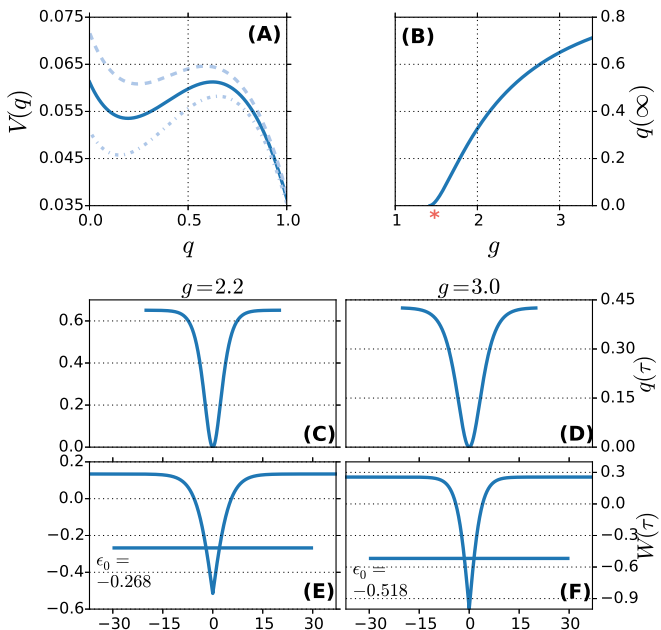}\protect\caption{\emph{Calculation of the largest Lyapunov exponent for a threshold-
linear network}\textbf{. (A)} Numerical integration of equation (\ref{eq:one_pop_EOM}).
The value of $x$ is determined through the requirement that the maximum
of the potential at nonzero $q$ equals its value at zero $q$. Compare
the form of the potential with the exact value of $x$ (solid line)
with those calculated with $x$ that deviates by $\pm1\%$ from the
correct value (dashed lines). \textbf{(B)} numerical solution for
the normalized variance $q(\infty)$as a function of $g$. \textbf{(C)-(F)}
The normalized autocorrelation function $q(\tau)$ found by integrating
the equation of motions (\ref{eq:one_pop_EOM}) using the correct
value of $x$ for two values of $g$ (top), and the corresponding
quantum potential $W=-\partial^{2}V/\partial q^{2}$ (bottom). The
ground state energies $\epsilon_{0}$ (horizontal lines) were found
numerically. In both cases they are negative implying a positive Lyapunov
exponent.\label{fig:SingPop-DMFT}}
\end{figure}

In the single population case, the evaluation of the Lyapunov exponent
is also relatively simple. In this case, the single-component Hamiltonian
is given by $\mathcal{H}=-\frac{\partial^{2}}{\partial\tau^{2}}+W(\tau)$
with a quantum potential $W=-\partial^{2}V/\partial\tdel^{2}$ where
$V$ is the classical potential (Fig \ref{fig:SingPop-DMFT}A) , which
can be easily evaluated once $\tdel(\tau)$ is known (Fig. \ref{fig:SingPop-DMFT}B).
Indeed, for the semi-linear network we find that the ground state
is negative for $g<\sqrt{2}$ and positive otherwise, implying that
$\lambda_{L}$ is negative for $g<\sqrt{2}$ and changes sign to a
positive value for $g>\sqrt{2}$ as expected in a chaotic state, see
Fig. \ref{fig:Phase-diagram}C-F. For the non linear case with$\nu=2$
the Lyapunov exponents are calculated near and above transition in
section \ref{sec:Criticality}.

\subsection{Numerical simulations }

\begin{figure}
\includegraphics[width=1\columnwidth]{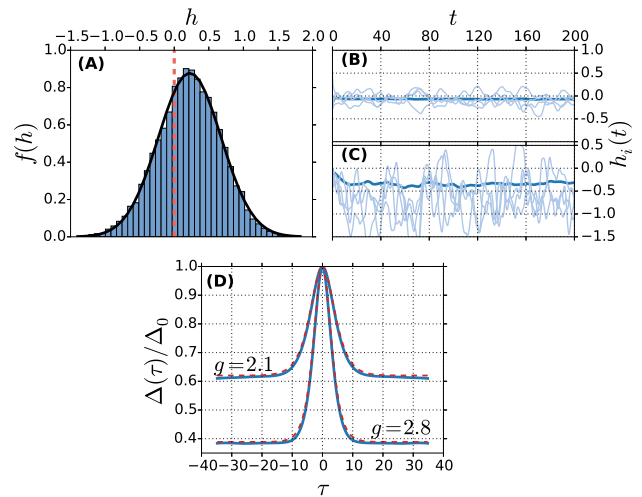}\protect\caption{\emph{Numerical simulations of single population with linear threshold
transfer function.} \textbf{(A)} Distribution of the local fields
$h^{i}$ in the fixed point regime. Thick curve shows normal distribution
given the theoretical mean and variance obtained from the mean field
theory. Dashed (red) vertical line marks the firing threshold which
is taken to be zero. \textbf{(B)},\textbf{(C)} activities of two networks
with gains $g=2.1$ and $g=2.8$ respectively. Bold lines show spatially
averaged local fields; thin lines are local fields of a sample of
four neurons. \textbf{(D)} the normalized AC function, $\Delta(\tau)/\Delta_{0}=1-\protect\tdel(\tau)$
for two values of gain parameters as in (B) and (C). Solid black lines
are the solutions of the equation of DMFT (\ref{eq:one_pop_EOM})
superimposed on simulation results (dashed lines). Simulations preformed
on a network ($N=6800$) with Gaussian distributed connections and
$\bar{g}=\sqrt{K}g$, where $K=680$ and external field $h^{0}=1$.
\textcolor{red}{\label{fig:one_pop_numerics}}}
\end{figure}

Numerical integration of the full network equations in (\ref{eq:one_pop_circuit_eq})
verifies our theoretical predictions. For threshold- linear network,
when $g<\sqrt{2}$ the network settles into a fixed point, with the
expected mean and variance of the local quenched fields (Fig. \ref{fig:one_pop_numerics}A).
When $g>\sqrt{2}$ chaos settles with temporal fluctuations that increase
with $g$. The population averaged currents remain almost constant
with small finite size fluctuations (Fig. \ref{fig:one_pop_numerics}B
and \ref{fig:one_pop_numerics}C). The chaotic behavior is characterized
first by the decay of the autocorrelation (Fig. \ref{fig:one_pop_numerics}D).
which agrees with the theoretical $\tdel(\tau)$, and second by a
positive LE. The latter was calculated from simulations using Wolf's
algorithm \cite{Wolf1985285}. The resultant values $\lambda_{L}=0.121$,
and $0.225$ for $g=2.2$ and $3.0$, respectively, agree well with
the values 0.126 and 0.232 obtained by numerically calculating the
ground state of the potential $W(\tau)$ (Fig. \ref{fig:Phase-diagram}D).
Simulations of a semi-quadratic network also verifies our analytical
results. Simulations results near and above the chaotic transition
are given below, in section \ref{sec:Criticality}.

\paragraph{Randomly diluted networks}

\begin{figure}
\includegraphics[width=1\columnwidth]{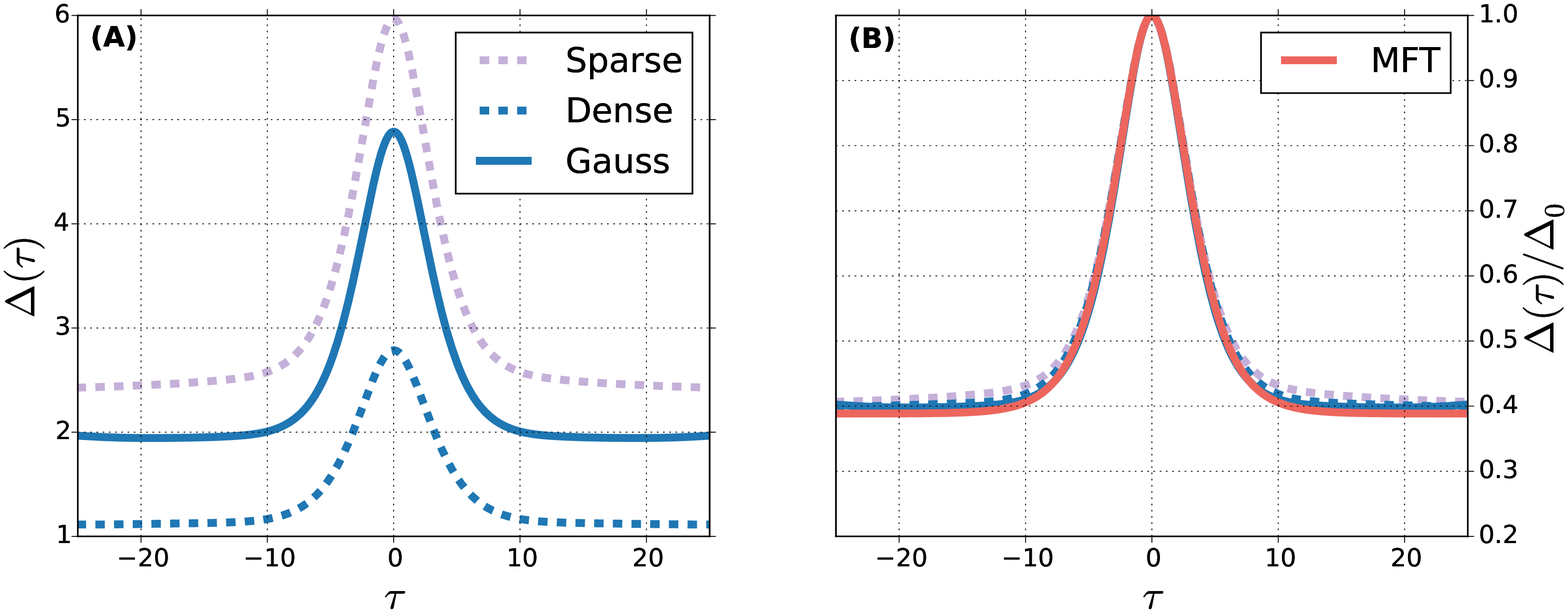}\protect\caption{\emph{Robustness to changes in the synaptic distribution}: \textbf{(A)}
Autocorrelation function for sparse ($p=0.05$), dense ($p=0.8$)
and gaussian connectivity Networks ($N=7000$, $h^{0}=1$). The gains
were chosen such that all networks have the same variance for the
connectivity distribution ($g=2.8$ for the Gaussian distribution
and $g\sqrt{1-p}=2.8$ for the randomly diluted networks).\textbf{
(B)} Normalized autocorrelation function $1-q(\tau)=\Delta(\tau)/\Delta_{0}$
compared with the solution of the DMFT. \label{fig:diff_K}}
\end{figure}

The above numerical results were for Gaussian distributed synaptic
connections. However, as shown above, in the limit of large mean number
of connections per neuron, $K$, the DMFT is expected to hold also
for randomly diluted networks, in which case, $g$ and $\bar{g}$
are related through (\ref{eq:gg0}) (Dashed lines in Fig \ref{fig:Phase-diagram}).
In the case of a single population with threshold linear transfer
function, comparing the behaviors of different connectivity schemes
(Gaussian, sparse and dense dilution) is relatively simple since the
normalized autocorrelation $\tdel(\tau)$ depends only on $g,$ not
on $\bar{g}$ (see above). These expectations are borne out by our
numerical simulations, shown in Fig. \ref{fig:diff_K}. For simulations
preformed on network with the same variance in their connectivity,
the calculated normalized autocorrelation $\tdel(\tau)$ is identical
in the Gaussian network and the randomly diluted networks with both
$p=K/N=0.05$ (sparse network) and $p=0.8$ (dense network). 

\section{Two populations with threshold-linear transfer function\label{sec:Two-populations-1}}

Here we address briefly the application of the general theory to the
case of a two-population network with one excitatory (denoted as $E$)
and one inhibitory ($I$) populations (Fig. \ref{fig:Schematics}C)..
For simplicity we study the example of the semi-linear transfer function.
Generalization to other values of $\nu$ is straight forward, as presented
in \ref{sec:DMFT}.

\subsection{Fixed point and its stability}

The fixed point equations for the variance and mean of the synaptic
current are:

\begin{equation}
\Delta_{k}=\sum_{l=E,I}g_{kl}^{2}\Delta_{l}\left\langle \left[z+x_{l}\right]_{+}\right\rangle _{z},\;k=E,I,
\end{equation}
\begin{equation}
m_{k}=\Delta_{k}^{\nu/2}\left\langle \left[z+x_{k}\right]_{+}\right\rangle _{z},\;k=E,I,
\end{equation}

alongside with the definition of $x_{k}$, 

\begin{equation}
x_{k}\sqrt{\Delta_{k}}=\sum_{l}\bar{g}_{kl}m_{l}+W_{k}m_{0}.
\end{equation}

The eigenvalues of the stability matrix $M_{kl}=g_{kl}^{2}H(-x_{l})$
are given by 

\begin{equation}
\Lambda_{\pm}=\frac{1}{2}\left[M_{EE}+M_{II}\pm\sqrt{\left(M_{EE}-M_{II}\right)^{2}+4M_{IE}M_{EI}}\right].\label{eq:TwoPop_eigenvalues-1-1}
\end{equation}
Note that the two eigenvalues are real. For a fixed point to be stable
$\Lambda_{+}<1$. Thus, the transition to chaos occurs at parameters
s.t. $\Lambda_{+}=1$. When this eigenvalue becomes larger than $1$,
one must solve the DMF equations for the chaotic state.

\subsection{Chaotic state}

The DMF equations can be written as, 

\begin{equation}
\left(1-\frac{\partial^{2}}{\partial\tau^{2}}\right)\tdel_{k}(\tau)=1-\tilde{g}_{kE}^{2}C_{E}(\tau)-\tilde{g}_{kI}^{2}C_{I}(\tau),\label{eq:DMFT}
\end{equation}

\begin{equation}
C_{k}(\tau)=\left\langle \left\langle \left[\sqrt{\tdel_{k}(\tau)}y+\sqrt{1-\tdel_{k}(\tau)}z+x_{k}\right]_{+}\right\rangle _{y}^{2}\right\rangle _{z},
\end{equation}
for $k\in\{E,I\}$, where

\begin{equation}
\tdel_{k}(\tau)=1-\Delta_{k}(\tau)/\Delta_{k}(0)
\end{equation}

are the normalized autocorrelation function for each population, and
here se set 
\[
\tilde{g}_{kl}\equiv g_{kl}\sqrt{\frac{\Delta_{l}(0)}{\Delta_{k}(0)}}.
\]

Unlike the simple case of a single population, no classical potential
function can be defined for the above two particle motion in (\ref{eq:DMFT}).
Nevertheless, the dynamical equations can be integrated numerically,
iteratively finding the values for $x_{k}$ that yield the desired
asymptotic behavior of the normalized variance, 

\begin{equation}
\tdel_{k}(\infty)=1-\Delta_{k}(\infty)/\Delta_{k}(0).
\end{equation}
Likewise, in general, the Hamiltonian governing the Lyapunov susceptibility
$\mathbf{G}(\tau)$ is not Hermitian. However, as stated above, we
expect the ground state to be real since the elements of $\mathbf{G}(\tau)$
are non negative by definition, and complex ground state would mean
oscillations around zero (See appendix \ref{sec:Appendix_Stability}).

\subsection{Numerical results}

Below we describe the chaotic state of this network, based on numerical
investigations. In these simulations, we focus on the balanced regime,
in which all the mean contribution of each population to the local
input is of order $\sqrt{K}$. In this case, to leading order, excitation
and inhibition cancel each other, and the mean activity of each population
is given by

\begin{equation}
m_{E}=-[J^{-1}W]_{E}m_{0}
\end{equation}
and

\begin{equation}
m_{I}=-[J^{-1}W]_{I}m_{0},
\end{equation}

where $W=[W_{E}W_{I}]^{T}$ is the vector of feedforward connections,
and $J$ is the 2x2 matrix of the mean recurrent connections, as defined
in the diluted model above. The balance conditions on the parameters
are

\[
\mathbf{\det J}<0,
\]
\[
J_{EI}W_{I}<J_{II}W_{E}
\]
and
\[
J_{IE}W_{E}<J_{EE}W_{I}.
\]

\begin{figure}
\includegraphics[width=1\columnwidth]{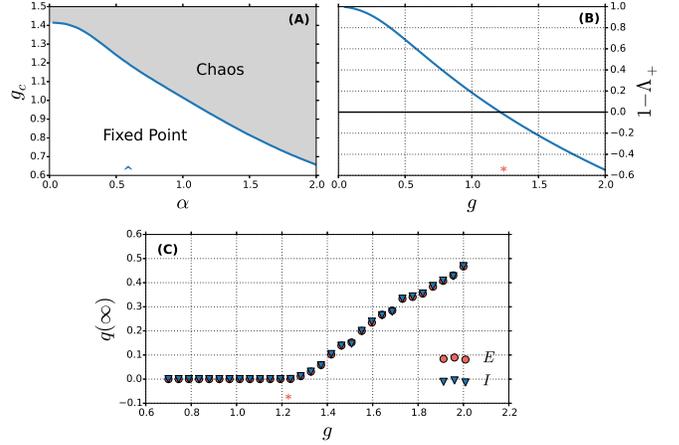}

\protect\caption{\label{fig:TP-phase_trans}\emph{Transition to chaos in a system of
two populations. }parameters used: $J_{EE}=J_{IE}=W_{E}=\alpha g$,
$J_{II}=-g$,$J_{EI}=-1.11g$ and $W_{I}=0.44g$. Note that $g$ is
a global gain multiplying all the synapses; $\alpha$ denotes the
excitatory efficacy.\textbf{ (A)} The critical value of $g$ as a
function of the excitatory efficacy $\alpha$. For $\alpha=0$ the
network is identical to the single inhibitory network and phase transition
occurs at $g=\sqrt{2}$. \textbf{(B)} Largest eigenvalue of the stability
matrix (\ref{eq:TwoPop_eigenvalues-1-1}) for a network with $\alpha=0.55$
{[}marked by '\textasciicircum{}' in panel (A){]}. MF predicts a phase
transition when $\Lambda_{+}=1$ (asterisks). \textbf{(C)} Normalized
variance, $\protect\tdel_{k}(\infty)$, for inhibitory (blue triangles)
and excitatory (red circles) populations as a function of the gain
calculated from network simulations with $\alpha=0.55$. Calculated
from simulation of a balanced diluted network with $N_{E}=N_{I}=3500$,
$K=700$, connectivity parameters as in panel (B), and external activity
$m_{0}=1$. Theory predicts a phase transition at $g=1.21$ (asterisks),
as seen in panel (B).}
\end{figure}

To demonstrate the properties of the two population system, we fix
all the values of the connections except for a global gain $g$, and
$\alpha$, which controls the excitatory connections. For $\alpha=0$
the network is reduced to the single inhibitory network with the parameters
used in section \ref{sec:Single-inhibitory-population}. In Fig. \ref{fig:TP-phase_trans}A
we show a section of the phase space showing the critical $g_{c}(\alpha)$
line. The critical curve is calculated by solving the eigenvalues
of the stability matrix (\ref{eq:TwoPop_eigenvalues-1-1}) for each
value of $\alpha$. Fig. \ref{fig:TP-phase_trans}B, shows an example
of the stability for $\alpha=0.55$. Numerical simulations confirm
the theoretical results as seen in Fig. \ref{fig:TP-phase_trans}C.
For convenience of comparison, parameters were chosen so that they
correspond to the same network parameters studied in \cite{Vreeswijk:1998uz,Renart:2010hj}.
Unlike the binary network in which no stable fixed point exists for
any $g,$ in the threshold-linear network a transition to chaos occurs
at $g=1.21$ (Fig \ref{fig:TP-phase_trans}B and C). Fig. \ref{fig:TP-phase_trans}C
indicates that the normalized variance of the two populations are
very similar to each other (see discussion in Section \ref{fig:TwoPop_critical}
below). 

\begin{figure}
\includegraphics[width=1\columnwidth]{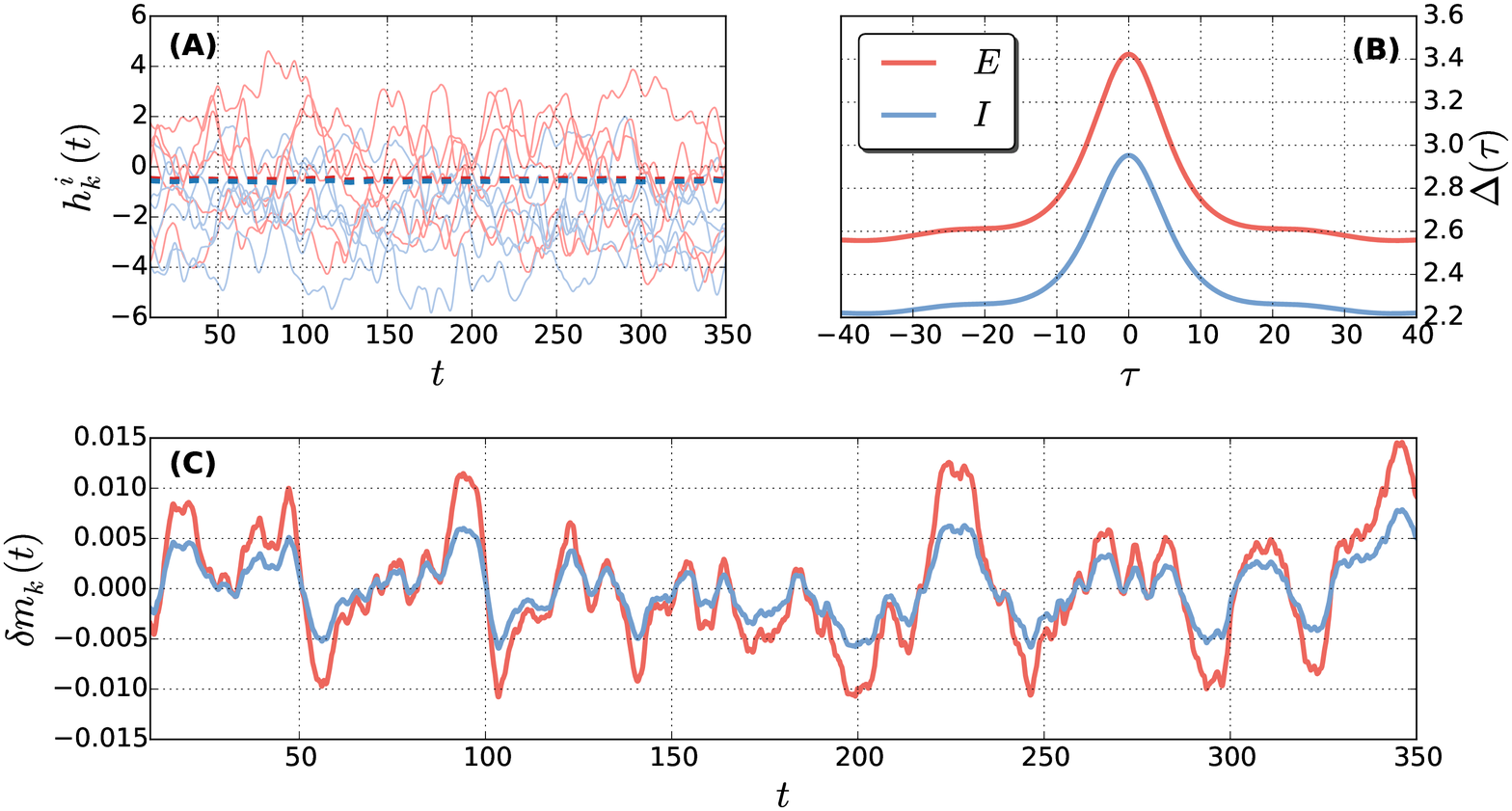}

\protect\caption{\label{fig:TP_chaotic}\emph{(color) Fluctuations of an E-I balanced
network with linear threshold transfer function}\textbf{. }Same parameter
set as in figure \ref{fig:TP-phase_trans}C. \textbf{(A)} Traces of
the local fields of six neurons from excitatory (red) and inhibitory
(blue) populations. Dashed lines show mean input into each population.\textbf{
(B)} Time-lagged autocorrelation function of two population computed
from the simulation. \textbf{(C)} Trace of the fluctuations in the
spatially averaged activities $\delta m_{k}(t)\equiv m_{k}(t)-\left\langle m_{k}(t)\right\rangle _{t}$
of both populations {[}same color codes as (A){]}. Simulations of
network similar the one used in Fig.\ref{fig:TP-phase_trans}B, with
$g=1.6$.}
\end{figure}

Figure \ref{fig:TP_chaotic} shows an example of the chaotic fluctuations
in the two population network. Same parameters were used as in Fig.
\ref{fig:TP-phase_trans}A and \ref{fig:TP-phase_trans}B with $g=1.6$,
within the chaotic phase.. As expected, inputs into neurons from both
populations show large fluctuations, while the mean activity of each
population is constant up to fluctuations of order $1/\sqrt{K}$ (Fig.
\ref{fig:TP_chaotic}A). The autocorrelation of both population decays
monotonically with $|\tau|$ to an equilibrium value (Fig. \ref{fig:TP_chaotic}B).
A signature of a dynamical balanced state is the substantial synchrony
in the fluctuations of the excitatory and inhibitory mean activities.
(Fig. \ref{fig:TP_chaotic}C). Consistent with Fig \ref{fig:TP-phase_trans}C
above, Fig. \ref{fig:TP_chaotic}B shows that the autocorrelation
functions of the two populations are nearly proportional to each other,
implying that the \emph{normalized} autocorrelation functions, $q_{E}(\tau)$
and $q_{I}(\tau)$ are approximately equal (as observed recently also
by \cite{aljadeff2014transition}; see Section \ref{sec:Criticality}
below). 

\section{Critical behavior at the onset of chaos\label{sec:Criticality}}

In this section, we analyze the characteristics of the system at the
onset of the chaotic state, and ask what features determine the critical
properties of this transition. As will be seen below, these properties
depend on the shape of the single neuron transfer function near the
origin. As such, the class of power law functions defined in \ref{eq:phi_nu}
can represent any continuous threshold function.

\paragraph{Absence of stable chaotic phase}

Before we explore the critical properties near the transition to chaos,
we note that not all transfer functions allow a stable chaotic solution.
For example, Eq (\ref{eq:dMFT_delta}) may not have a solution satisfying
all boundary conditions. In these instances dynamic is either stationary
or explosive (i.e., 'epileptic'). An example of such behavior is the
exponential curve $\phi(x)=e^{x}$, as shown in Appendix \ref{sec:Exponential-transfer-function}.
On the other hand all $\phi(x)$ which saturate as $x\rightarrow\infty$
are expected to exhibit a stable chaotic state for large values of
$g$.

In the following, we focus our study on transfer function of the type
(\ref{eq:phi_nu}), with $\nu>1/2$, which exhibits a phase transition
from FP to stable chaos.

\subsection{Critical properties: A single population}

When $\nu>1/2$, the critical properties near the transition to chaos
depends on the value of $\nu$. To see this, we first consider the
case of a single inhibitory population. The critical value of $g$
is given by the value at which $g^{2}\langle[\phi'(\sqrt{\Delta}z+u)]^{2}\rangle=1$,
see Eq (\ref{eq:SP_FP_sol}). As the chaotic state approaches the
critical $g$ from above, the amplitude of the time dependent fluctuations
becomes small, hence, we can expand the MF equations in powers of
$\tdel(\tau)$ (which is small at all $\tau$). In addition, we will
expand the equations in the small static parameter $\delta x\equiv x-x_{c}$
and the bifurcation parameter $\epsilon\equiv g^{2}/g_{c}^{2}$-1. 

To leading order, the DMF equation (\ref{eq:DMFT}) takes the form
(see appendix\ref{sec:Appendix_Rescaling}) 
\begin{equation}
\frac{\partial^{2}\tdel(\tau)}{\partial\tau^{2}}=a\left(\epsilon\right)+b\epsilon\tdel(\tau)+c\tdel^{\rho}(\tau),\label{eq:phi_nu_dynamic}
\end{equation}
where $b$ and $c$ are parameters of order unity, and the exponent
$\rho$ obtained from expansion of the firing rate autocorrelation,
$C(\tau)$ (see appendix \ref{sec:Appendix_Rescaling}) is 

\begin{equation}
\rho=\begin{cases}
\frac{3}{2}, & \;\frac{1}{2}<\nu\leq\frac{3}{2}\\
2, & \;\frac{3}{2}<\nu
\end{cases}.
\end{equation}

The constant term $a$ vanishes at the transition and depends on $\epsilon$.
In order for a nontrivial solution to exist, all the terms in (\ref{eq:phi_nu_dynamic})
should be of the same order of $\epsilon,$ hence the time scale of
$\tdel(\tau$) should scale as $\tau_{eff}\sim1/\sqrt{\epsilon}$.
The amplitude of $\tdel(\tau)$ should scale as $\epsilon^{2}$ for
$\nu\leq3/2$ and as $\epsilon$ for larger $\nu$. Finally, $a$
should scale as $O(\epsilon\tdel)$ (as indeed found by an explicit
evaluation of $a$, see appendix), yielding the following scaling
behavior,

\begin{equation}
\tdel(\tau)=\epsilon^{\frac{1}{\rho-1}}f\left(\frac{\tau}{\sqrt{\epsilon}}\right),\label{eq:scaling_del}
\end{equation}
where $f(x)$ is of order $1$. The Hamiltonian in (\ref{eq:Hamiltonian})
scales as 
\begin{equation}
W(\tau)=\epsilon F\left(\frac{\tau}{\sqrt{\epsilon}}\right),\label{eq:Hamiltonian_rescaled}
\end{equation}
where $F(x)$ is some other function of order $1$. From (\ref{eq:Hamiltonian_rescaled})
it implies that the largest Lyapunov exponent, Eq (\ref{eq:Lambda_max}),
scales as 

\begin{equation}
\lambda_{L}=O(\epsilon).\label{eq:scaleLambda}
\end{equation}

\begin{figure}
\includegraphics[width=1\columnwidth]{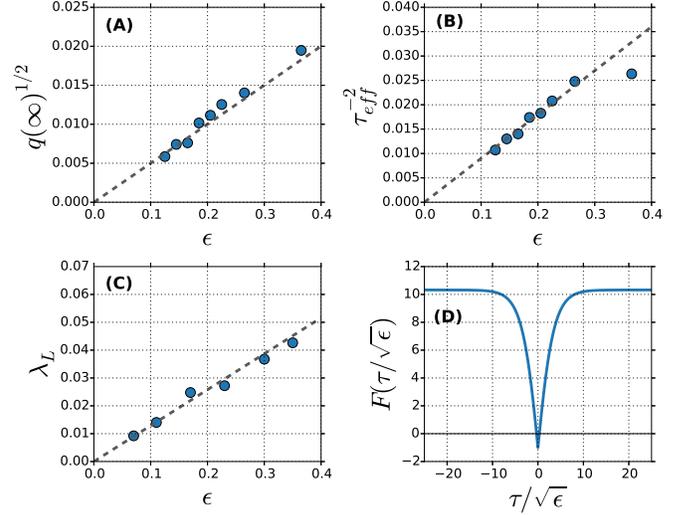}\protect\caption{\textbf{\label{fig:Near-Phase-Transition-nu1}}\emph{Critical behavior
of a single inhibitory population with linear transfer function.}
Normalized variance \textbf{(A)}, network relaxation time \textbf{(B)},
and largest Lyapunov exponent \textbf{(C)} as a function of the distance
from the critical point $\epsilon=\frac{1}{2}g^{2}-1$. Circles show
average over 20 simulations (Gaussian connectivity, $N=6000$, $h^{0}=1$)
and dashed lines show theoretical predictions (See Appendix \ref{sec:Appendix_Rescaling}).
\textbf{(D)} Rescaled 1D quantum potential $F(\tau/\sqrt{\epsilon)})$,
Eq (\ref{eq:Hamiltonian_rescaled}), for the Hamiltonian in (\ref{eq:Hamiltonian}). }
\end{figure}

\begin{figure}
\includegraphics[width=1\columnwidth]{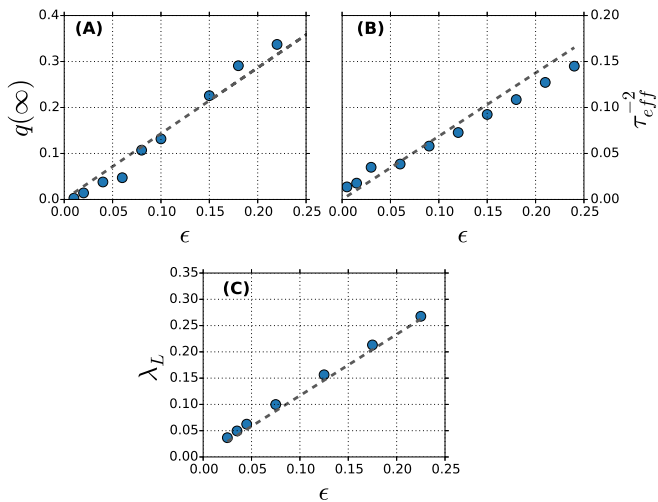}\protect\caption{\label{fig:Near-Phase-Transition-nu2}\emph{Critical behavior of a
single inhibitory population with quadratic ($\nu=2)$ transfer function}.
Normalized variance \textbf{(A)}, network relaxation time \textbf{(B)}
and LE \textbf{(C)} for small $\epsilon\equiv g^{2}/g_{c}^{2}-1$
gain above the critical point. Circles show average over 20 simulations
(Gaussian connectivity in the balance regime, $N=6000$, $K=1200$,
$h^{0}=1$) and dashed lines show theoretical predictions (See Appendix
\ref{sec:Appendix_Rescaling}).}
\end{figure}

We have confirmed the above predictions numerically, for the cases
of $\nu=1$ (Fig. \ref{fig:Near-Phase-Transition-nu1} ) and $\nu=2$
(Fig. \ref{fig:Near-Phase-Transition-nu2}). 

In this work, we assume a synaptic transfer function of the form (\ref{eq:phi_nu}).
In \cite{Sompolinsky:1988ta} the variance of a network with a sigmoid
synaptic transfer function, $\phi(h)=\tanh(h)$, and $h^{0}=0$, was
shown to scale as $\Delta(\tau)=\epsilon f\left(\frac{\tau}{\sqrt{\epsilon}}\right)$
and its LE as $\lambda_{L}=O(\epsilon^{2})$. This behavior results
from the $h\rightarrow-h$ symmetry of the system, which implies that
$\Delta_{0}$ vanishes at the transition. This symmetry is not expected
to hold in biological networks. Thus, the scaling shown in (\ref{eq:scaling_del})
and (\ref{eq:scaleLambda}), in which the LE is larger than in the
symmetric case, reflects the behavior in generic neuronal networks.

In \cite{Wainrib:2013is} the authors study the distribution of equilibria
points above criticality, for a single population with sigmoidal transfer
function. An interesting result is that the mean (with respect to
realizations of $J_{ij}$) number of equilibria behaves just like
the Lyapunov exponent. Consequently, (\ref{eq:scaleLambda}) may also
elucidate the topological complexity of the flow above criticality
for threshold power law transfer functions.

\subsection{Multiple populations}

\begin{figure}
\includegraphics[width=1\columnwidth]{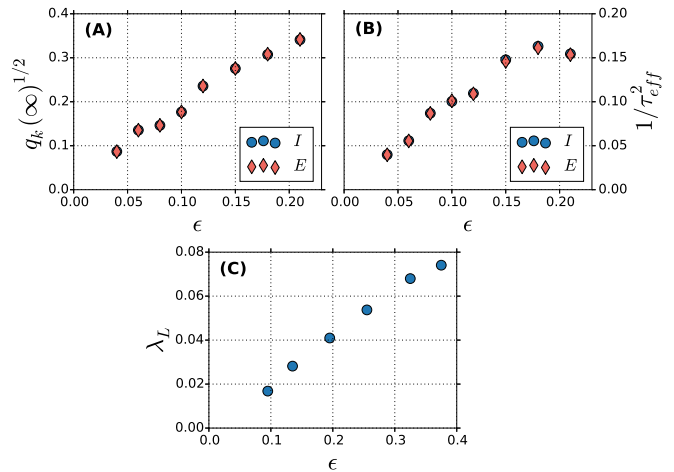}

\protect\caption{\label{fig:TwoPop_critical}\emph{Critical behavior of a network with
two populations and a linear transfer function.} Diamond (red) and
circles (blue) denote excitatory and inhibitory populations respectively.
\textbf{(A)} The normalized variance of the two population, $\protect\tdel_{E}(\infty)$
and $\protect\tdel_{I}(\infty)$, show the same critical behavior
(a linear rise from zero) as the single population (Eq (\ref{eq:scaling_del})).
Due to the linearity, the normalized autocorrelations are nearly equal
to each other near the critical point (see appendix \ref{sec:Appendix_Rescaling}).
\textbf{(B)} Network relaxation times of both populations are similar
at the scaling limit. \textbf{(C)} LE of the entire network calculated
over all populations, Eq (\ref{eq:LE_theory}), scale as a single
population, Eq (\ref{eq:scaleLambda}). Simulations conducted on network
with diluted connectivity matrix ($N=3600$, $K=600$ for each population),
and parameter set as in Figs. \ref{fig:TP_chaotic} and \ref{fig:TP-phase_trans}C.}
\end{figure}

The above critical properties were derived for the case of a single
population. However, we argue that these properties hold also for
multiple populations. In the case of several populations, transition
to chaos occurs when the largest eigenvalue of the stability matrix,
$\mathbf{M}$, (\ref{eq:stabFP}), equals $1$. Near the transition
(on the chaotic side), this eigenvalue is slightly larger than $1$
while the real part of the rest remain stable. To leading order, the
unstable \emph{direction}, $R^{1},$ in the space of $\tdel_{k}(\tau)=1-\Delta_{k}(\tau)/\Delta_{k}(0)$
remains the same as the marginally stable eigenvector at the critical
point. Hence, near the transition the dominant direction of temporal
fluctuations are along the critical eigenvector, and the dynamic equations
collapse into one dimension, similar to the one population case. In
general, we expect that all $\tdel_{k}$ have nonzero components on
the critical direction, hence $q_{k}(\tau)\approx q(\tau)R^{1}$ and
the critical properties are the same as the single population properties,
above, see Appendix \ref{sec:Appendix_Rescaling}. Similar arguments
apply to the scaling properties of the quantum Hamiltonian, hence
the LE scales as (\ref{eq:scaleLambda}). These results are supported
by simulations of an excitatory - inhibitory network, defined in Section
\ref{sec:Two-populations-1}. The simulations displayed in Fig, \ref{fig:TwoPop_critical}
demonstrate the universality of the critical properties of the transition
to chaos. In the specific case of a threshold-linear model, we find
for a wide range of parameters, $R^{1}\approx(1,1,1,...\text{1})$,
implying that near the transition, the normalized autocorrelations
of all populations are not only proportional but are expected to be
approximately equal to each other, as detailed in appendix \ref{sec:Appendix_Rescaling}. 

Note that the one-dimensional character of the chaotic fluctuations
is exact only asymptotically close to the transition. Away from the
transition, the non-linearity of $C(\tdel)$ couples the unstable
mode to the other modes, inducing fluctuations in all modes. Interestingly,
we find numerically that in many cases, even far from the transition,
the autocorrelations are still close to being equal, $\tdel_{k}(\tau)\approx\tdel(\tau)$
as is apparent in the example of Fig. \ref{fig:TP-phase_trans}C and
\ref{fig:TP_chaotic}B. Similar observations have been made recently
in \cite{aljadeff2014transition}. 

\section{Transition to chaos in spiking networks\label{sec:Spiking}}

In this section, we inquire under what conditions the transition to
chaos observed in the rate dynamics occurs also in spiking networks.
In contrast to rate models, the membrane potentials of spiking networks
are not at a fixed point as long as some of the neurons are active.
Hence, in general, there is no transition from fixed point to chaos.
An exception is the case of networks with slow synapses. If the synaptic
time constant is long compared to the spiking dynamics, it is possible
that for low synaptic strength, the spiking dynamics is averaged out
at the level of the slow synaptic currents, which will therefore stay
approximately constant. Thus, there might be a critical $g$ in which
the state with almost constant synaptic currents undergoes an instability
leading to a chaotic (or at least temporally irregular) state in which
temporal fluctuations of the synaptic currents are large even at the
scale of the synaptic time constant. 

To explore the possibility of transition to chaos in spiking networks
with slow synaptic time constant, we consider a network of spiking
neurons that fire with inhomogeneous Poisson statistics, specified
by the instantaneous rate of each neuron, $r^{i}(t)=\phi\left(h^{i}(t)\right)$.
We assume that the network is in an asynchronous state \cite{Hansel:1996he}
and that the typical firing rate is large compared with the inverse
synaptic time constant,$\tau_{s}^{-1}$.

Focusing on a single inhibitory populations, we write the dynamic
equation of the local fields as, 
\begin{equation}
\tau_{s}\frac{d}{dt}h^{i}(t)=-h^{i}(t)+\sum_{j=1}^{N}\mathcal{J}^{ij}\xi^{j}(t)+\tau_{0}\bar{g}r(t)+h^{0}.\label{eq:Spiking-model}
\end{equation}
where $\mathcal{J}^{ij}$ is either Gaussian distributed connections
or the corresponding randomly dilute network, as appears in Eq (\ref{eq:one_pop_circuit_eq})
above. The spike train $\xi^{j}(t)=\sum_{k}\delta\left(t-t_{k}^{j}\right)$
of the presynaptic neuron $j$ at times $t_{k}^{i}$ represents the
random Poisson process of neuron $j$. The non-linear rate function
is given by the rectified linear transfer function $\phi(x)=\tau_{0}^{-1}\left[x\right]_{+}$where
$\tau_{0}$ is a microscopic time constant, which is related to the
inverse slope of the single neuron $f-I$ curve. Note that in this
model, the connections have units of time, hence we define $g^{2}=N\langle(\mathcal{J}^{ij})^{2}\rangle\tau_{0}^{-2}$
so that $g$ (and $\bar{g}$) are dimensionless. For simplicity of
notation, in the following we measure rates and times in units of
$\tau_{0}$, i.e., we use units such that $\tau_{0}=1$.

To understand the effect of spiking dynamics we can separate the synaptic
input into a rate contribution and a spiking noise contribution

\begin{equation}
\eta_{i}(t)=\eta_{i}^{r}(t)+\eta_{i}^{sp}(t),\label{eq:spiking_nu}
\end{equation}
where
\begin{equation}
\eta_{i}^{r}(t)=\sum_{j=1}^{N}\mathcal{J}^{ij}\phi(h_{j}(t))
\end{equation}
\begin{equation}
\eta_{i}^{sp}(t)=\sum_{j=1}^{N}\mathcal{J}^{ij}(\xi^{j}(t)-\phi(h_{j}(t))).
\end{equation}

The auto-covariances of the two term on the RHS of (\ref{eq:spiking_nu})
are 
\begin{equation}
\langle\eta_{i}^{r}(t)\eta_{i}^{r}(t+\tau)\rangle=g^{2}C(\tau),
\end{equation}
and
\begin{equation}
\langle\eta_{i}^{sp}(t)\eta_{i}^{sp}(t+\tau)\rangle=g^{2}r\delta(\tau),
\end{equation}
where $C(\tau)=\left\langle \phi_{i}(t)\phi_{i}(t+\tau)\right\rangle $
is the autocorrelation of the rate functions given by (\ref{eq:dMFT_C})
(in units of $\tau_{0}^{-2}$), and$\delta(\tau)$ is the Dirac delta
function. The last equation results from the average over the Poisson
process, $\langle(\xi^{j}(t)-\phi(h_{j}(t))(\xi^{j}(t+\tau)-\phi(h_{j}(t+\tau))\rangle=\phi(h_{j}(t))\delta(\tau)$
. Thus, the Poisson noise is equivalent to an additive white noise
with amplitude $g^{2}r$, where $r$ is the population averaged rate
$r$ (in units of $\tau_{0}^{-1}$). Thus, the spiking noise can be
incorporated into the dynamic mean field theory, yielding,

\begin{equation}
\left(1-\tau_{s}^{2}\frac{\partial^{2}}{\partial\tau^{2}}\right)\Delta(\tau)=g^{2}C(\tau)+g^{2}r\delta(\tau),\label{eq:poisson_delta_dynamics-1}
\end{equation}

\subsection{Perturbation analysis of the spiking noise}

As explained above, we are interested in the regime of large $\tau_{s}$
(which is the synaptic time constant relative to $\tau_{0}$). This
limit can be illuminated by writing the rescaled mean field dynamics,
\begin{equation}
\left(1-\frac{\partial^{2}}{\partial t^{2}}\right)\Delta(\tau)=g^{2}C(t)+\frac{g^{2}r}{\tau_{s}}\delta(t),\label{eq:poisson_delta_dynamics}
\end{equation}
where we have scaled time so that $t=\tau/\tau_{s}$. The above equation
demonstrates that the contribution of the Poisson noise (proportional
to $\frac{g^{2}r}{\tau_{s}}$ ) is small relative to the rate contributions
(which are proportional to $g^{2}r^{2}$ ) and the ratio between the
two is of the order of $(r\tau_{s})^{-1}$. 

\begin{figure}
\includegraphics[width=1\columnwidth]{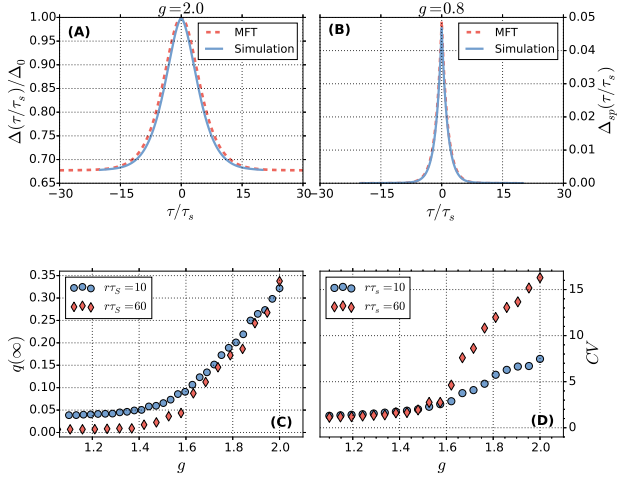}\protect\caption{\label{fig:Spiking-AC}\emph{Network of spiking neurons with inhomogeneous
Poisson statistics. }Simulation results for a network that follows
the dynamic equation in \textbf{(\ref{eq:Spiking-model})}.\textbf{
(A)} Above the critical point of the underlying rate dynamics, $g_{c}=\sqrt{2}$,
the rate term dominates the DMF equation, (\ref{eq:poisson_delta_dynamics})
and the autocorrelation calculated from simulation ($r\tau_{s}=30$
,solid line) is fully explained by autocorrelations of the rates,
found by (\ref{eq:one_pop_EOM}) (dashed line). \textbf{(B)} For values
of $g$ well below the rate transition, the rates do not have intrinsic
fluctuations, and the fluctuations in the fields is solely due to
the Poisson firing. MFT results of (\ref{eq:poisson_delta_sol}) (dashed
line) agrees with simulations ($r\tau_{s}=30$, sold line). In intermediate
values of $g$, rate and spike fluctuations interact nonlinearly.
\textbf{(C) }Normalized variance of the local fields $q(\infty)$
from simulations of networks at different gain levels in the transition
region. For higher levels of $r\tau_{s}$, the crossover between variance
dominated by the Poisson process and variance dominated by the fields
dynamics becomes sharper. \textbf{(D)} The coefficient of variation
($CV$), Eq (\ref{eq:CV_def}), for different values of gain. For
low values of $g$, the $CV$ approaches $1$, indicating a Poisson
process with (almost) constant rates. For higher values of the gain,
the $CV$ is larger, as expected from an inhomogeneous Poisson process.
The crossover between the two regimes becomes sharper as $r\tau_{s}$
increases. Simulations preformed using an inhibitory population with
Gaussian distributed connectivity ($N=3500$, $K=700$) and $\bar{g}=h^{0}=1$.}
\end{figure}

For gain values above $g_{c}=\sqrt{2}$, and $r\tau_{s}\gg1$, the
noise contribution will be small compared to the unperturbed rate
autocorrelation given by the solution of (\ref{eq:one_pop_EOM}) (Figure
\ref{fig:Spiking-AC}A). Below $g_{c}$, the only time dependence
comes from the spikes. To study this regime, we consider the spikes
contribution as a small perturbation around the static autocorrelation,
$\Delta(t)=\Delta_{st}+\Delta_{sp}(t)$, where $\Delta_{st}$ is the
static solution for $\Delta$ in the rate model, and expand the equation
above to linear order in $\Delta_{sp}(t)$ yields 

\begin{equation}
\left(1-\frac{\partial^{2}}{\partial t^{2}}\right)\Delta_{sp}(t)=g^{2}C'\Delta_{sp}(t)+\frac{g^{2}r}{\tau_{s}}\delta(t),\label{eq:linear}
\end{equation}
where $C'=H(-x)$, see Eq (\ref{eq:SP_FP_stab}). Solving (\ref{eq:linear})
yields 
\begin{equation}
\Delta_{sp}(\tau)=\frac{g^{2}r}{2\lambda\tau_{s}}e^{-\lambda\frac{\left|\tau\right|}{\tau_{s}}},\label{eq:poisson_delta_sol}
\end{equation}
where, following (\ref{eq:SP_FP_stab}),
\begin{equation}
\lambda^{2}\equiv1-g^{2}H(-x).
\end{equation}
Here $x$ is the unperturbed net input, Eq (\ref{eq:x_def}), and
we have substituted back $\tau=t\tau_{S}$. {[}Note that $g$ is in
units of $\tau_{0}${]}. Figure \ref{fig:Spiking-AC}B displays the
simulation results for the autocorrelation $\delta\Delta(\tau)=\Delta(t)-\Delta(\infty)$
for $g$ below the critical value, and the theoretical estimates $\Delta_{sp}(\tau)$. 

\subsection{Scaling analysis}

Approaching the transition point from below, the Poisson contribution,
(\ref{eq:poisson_delta_sol}), is amplified by the divergent response
of the fields' dynamics as $\lambda\to0$. For a finite $r\tau_{s}$
however, the autocorrelation remains finite at all $g$ and the transition
is smoothed. Indeed, Figure \ref{fig:Spiking-AC}C shows that $\tdel(\infty)$
increases smoothly as a function of $g$ but its increase become sharper
the larger $r\tau_{s}$is is. Figure \ref{fig:Spiking-AC}D shows
similar behavior for a measure of the fluctuations in the spike times,
known as Coefficient of Variation, $CV$, defined as the ratio between
the standard deviation of the ISI distribution, $\sigma_{ISI}$, and
its mean, $\mu_{ISI}$ \cite{Greengard:2001fv},boundaty
\begin{equation}
CV=\frac{\sigma_{ISI}}{\mu_{ISI}}.\label{eq:CV_def}
\end{equation}
 The $CV$ increases from $CV\approx1,$ at small $g$, which the
value for a homogeneous Poisson values, to substantially larger values
above $g=\sqrt{2}$, due to the fluctuations in the underlying rates.
This increase is smooth but become sharper for large values of $r\tau_{s}$. 

To study the effect of the small spiking noise on the transition,
one needs to perform a nonlinear analysis. Here we use a scaling analysis,
similar to that of second order phase transition in equilibrium statistical
mechanics. The scaling regime is characterize by two variables, $r\tau_{s}\gg1$
and $\epsilon\equiv g^{2}/g_{c}^{2}-1,$ $|\epsilon|\ll1$. We write
the normalized variance near the transition as
\begin{equation}
\tdel(\infty)=\frac{1}{(r\tau_{s})^{\alpha}}f(z),\label{eq:spiking_delta_scale}
\end{equation}
where the scaling variable $z$ is given by

\begin{equation}
z=r\tau_{s}|\epsilon|^{\beta}sign(\epsilon).
\end{equation}
Far below the transition, $z\to-\infty$ and according to (\ref{eq:poisson_delta_sol})
$\tdel^{-}(\infty)\sim\frac{1}{(r\tau_{s})\lambda}\sim\frac{1}{(r\tau_{s})\epsilon^{1/2}}$,
requiring $f(z\rightarrow-\infty)\propto z^{\alpha-1}$ and $\beta(1-\alpha)=\frac{1}{2}$.
Similarly, above transition $z\to\infty$, and from (\ref{eq:scaling_del})
$\tdel^{+}(\infty)\sim\epsilon^{2}$ entailing $f(z\rightarrow\infty)\propto z^{\alpha}$
and $\alpha\beta=2$. It follows that the critical exponents are 
\begin{equation}
\alpha=\frac{4}{5},\;\beta=\frac{5}{2}.
\end{equation}

The behavior of the effective relaxation time of the network can also
be treated in a similar manner. In the absence of the spiking noise,
the effective time constant of the autocorrelation diverges as the
transition is approached from above, as $\tau_{eff}\sim\epsilon^{-1/2}$
(Eq (\ref{eq:scaling_del}) and Fig. \ref{fig:Near-Phase-Transition-nu1}B).
In the presence of small spiking noise this time constant never diverges
and we write, 
\begin{equation}
\tau_{eff}^{-1}=\frac{1}{(r\tau_{s})^{\gamma}}F\left(z\right),\label{eq:spiking_tau_scale}
\end{equation}
where $\tau_{eff}$ is the effective correlation time in units of
$\tau_{s}$. Here, the critical behavior both below and above transition,
$z\to\pm\infty$, is similar (as can be seen from (\ref{eq:poisson_delta_sol}))
and $\tau_{eff}^{-1}\sim\lambda\sim\epsilon^{1/2}$ implying $F(z\to\pm\infty)\propto|z|^{\gamma}$
and $\beta\gamma=\frac{1}{2}$, or 
\begin{equation}
\gamma=\frac{1}{5}.
\end{equation}

\begin{figure}
\includegraphics[width=1\columnwidth]{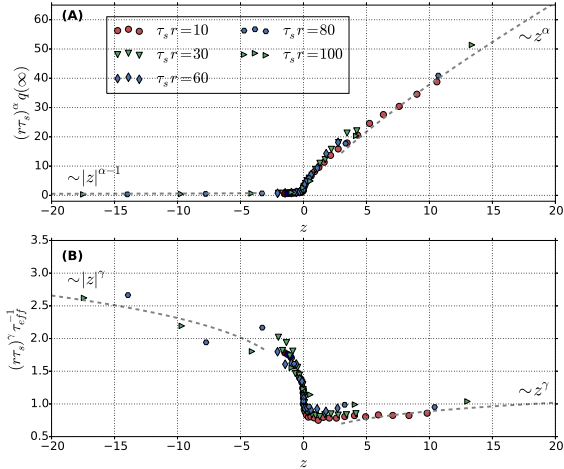}\protect\caption{\emph{Scaling behavior near criticality. }Rescaled form for the normalized
variance, $(r\tau_{s})^{\alpha}\protect\tdel(\infty)$, \textbf{(A),
}and inverse effective relaxation time $(r\tau_{s})^{\gamma}\tau_{eff}^{-1}$,
\textbf{(B),} using the scaling variable $z=\tau_{s}r|\epsilon|^{\beta}sign(\epsilon)$.
The data reveals the underlying scaling functions $f(z)$ and $F(z)$,
respectively, consistent with (\ref{eq:spiking_delta_scale}) and
(\ref{eq:spiking_tau_scale}) Dashed lines show the predicted asymptotic
behavior (see text). Simulations preformed on Gaussian network of
various sizes ($N=2000$, $3500$ and $6000$, $h^{0}=1$) and with
various synaptic time scales (see legend). Effective relaxation was
calculated from simulations by taking $\tau_{eff}=\int\tau\delta\Delta(\tau)d\tau/\int\delta\Delta(\tau)d\tau$
for $\tau\geq0$. \label{fig:Spiking-scaling}}
\end{figure}

Note that the above results predict that at the (rate dynamic) transition
($\epsilon=0)$ the amplitude of the variance vanishes as $(r\tau_{s})^{-4/5}$
and the effective time constant diverges $(r\tau_{s})^{1/5}$, respectively.
Simulation results that support these analytical predications are
presented in Fig. \ref{fig:Spiking-scaling}.

\section{Discussion\label{sec:Discussion}}

\paragraph{Universality of the transition to chaos across network architectures}

The results presented here show the universality of the dynamical
transition from a fixed point to chaos in large networks across different
network architectures. The theory, as well as simulations, show no
dependence on the detailed distribution of the synaptic connectivity
beyond the first two moments. In particular, a Gaussian network behaves
similarly to randomly dilute networks with either sparse or dense
connectivity. The two architectures differ in the mechanism by which
fluctuations are not suppressed by the high connectivity. In the Gaussian
networks, the mean of the connection distribution is assumed to be
smaller by a factor of $1/\sqrt{N}$ than their variance. In contrast,
in the dilute networks, the mean inputs from each population is strong
compared to the fluctuations and the fluctuations are nevertheless
potent due to the dynamic balance of excitation and inhibition. This
balancing amplifies not only the temporal fluctuations but also the
time averaged ones. For instance, the neurons in the dilute networks
exhibit a broad distribution of time averaged firing rates. In the
threshold linear case, the distribution of the firing rates will have
a truncated Gaussian shape with a width which is related to $\Delta_{k}(\infty).$
This balance mechanism is similar to the previously shown balanced
state in spiking or binary networks. Here we show that this process
is quite general and holds also in smooth dynamics with graded nonlinearity.
It should be also noted that most of the previous work on balanced
states assumed sparse connectivity, i.e., $K\ll N$. In this limit,
neurons share only few common sources, hence the network state is
naturally asynchronous. Interestingly, we show here that even dense
networks (where e.g. $K/N$ is as large as $0.8$, see Fig.\ref{fig:diff_K})
exhibit the same transition from a fixed point to asynchronous chaos,
as in sparse or Gaussian networks (see also \cite{Renart:2010hj,Helias:2014by}).
It would be interesting to prove analytically the absence of stable
synchronous states in these systems. 

The analytical and numerical investigations indicate that the transitions
to chaos in a single population and a two populations (E-I) network
is of the same nature and the DMFT predicts that this is true also
for networks with a general multiple populations architecture. The
extension of the DMFT to multiple populations is similar in spirit
to previous work on stochastic multi-population networks with weak
connections (of order $1/K$ ) \cite{Ginzburg:1994wm} and remains
valid as long as the number of populations is small compared to $K$.
The structure of the DMFT in the case of multi-population is different
than in the single population case in that the dynamics of the autocorrelation
function is not governed by a potential; thus, numerically solving
these equations is more challenging. Furthermore, although near the
transition the autocorrelations behave qualitatively similar to the
single population case, deep in the chaotic phase, depending on the
parameter value, they may not be monotonically decreasing with time
but might exhibit damped oscillations. Likewise, in multiple population
networks, the 'quantum' operator for the Lyapunov exponent is non
Hermitian, the implications of which needs to be further explored.

\paragraph{Stability against uniform perturbation}

Finally, we have focused primarily on asynchronous dynamical states
and instabilities driven by the random components of the interactions.
However, in the general architectures considered here, the fixed point
state may undergo first an instability associated with the population
averaged perturbations. Interestingly, due to the inherent nonlinearity
of our system, the uniform and local degrees of freedom are coupled.
As a result, the response to uniform perturbations involves equations
that couple the susceptibility of the population mean activities and
that of the population activity variances, Eq.(\ref{eq:uniformstabeq}).
This, together with (\ref{eq:stabFP}) constitutes the stability conditions
for the fixed points of our general architecture. These results can
also be interpreted as results regarding the eigenvalue spectrum of
the random matrix of effective couplings which incorporate the local
gains $\phi'(h_{l}^{i})$ from linearizing around the fixed points,
which are correlated with the random coupling (see also \cite{Ahmadian:2015co}). 

In the examples studied here, the uniform instability leads to a runaway
of the dynamics. In general, in the multi-population case, the fixed
point instability to uniform perturbations may exhibit an instability
leading to coherent oscillatory states, and these states may further
become destabilized by the random connectivity yielding possibly chaotic
states in the form of partially synchronous oscillations. The exploration
of these states is beyond the scope of this paper. Finally, we emphasize
that in the case of dilute networks, the primary focus of our work,
the underlying strong inhibition puts the system aways from these
instabilities as long as they are in states with excitation-inhibition
balance. In particular, multiply stable states occur only under special
conditions \cite{vanVreeswijk:2005uo,LitwinKumar:2012go}.

\paragraph{\emph{Spatially modulated networks}}

In the multi-population model, the functional structure of the entire
network is given in terms of pairwise connectivity which depends on
the subpopulation membership of the pair. In the brain, connectivity
probability depends also on the distance between neurons. This distance
can be either with respect to physical location or functional, namely
the preferred stimulus of a neuron. Likewise, the external input to
neurons is not in general homogeneous but may depend on the physical
or functional coordinate of each neuron. Because these location dependencies
are broad and smooth, the system can be well described by dividing
it into columns, each represents many neurons having similar location.
The total system, a hyper-column can thus be considered a special
case of our general multi-population architecture. In the large $N$
limit, the sums over columns will turn into smooth integrals over
the spatial coordinates. In general, the architecture of a neural
circuit if large enough will consist of both genuinely discrete subpopulations
(e.g., Excitatory and Inhibitory) as well as continuously varying
coordinates. An example is a model of orientation hyper-column in
V1 where the connectivity between neurons depend on the difference
in their preferred orientation and the external input depends on the
difference between the preferred orientation and the stimulus orientation
\cite{vanVreeswijk:2005uo}. In this case, the balance equations (\ref{eq:balance})
determine the tuning properties of the mean firing rates of the network.
The DMFT describes the statistical properties of the spatiotemporal
fluctuations in the network dynamics.

\paragraph{The shape of the nonlinear transfer function}

We have found that the existence of a transition from a fixed point
to chaos as well as its critical properties when it exists, depend
on the shape of the input-output transfer function near the origin
(i.e., near the firing threshold). An important result is that for
a transfer function rising as square root or sharper, i.e., for $\phi(x)\propto x^{\nu}$with
$\nu\leq1/2$, there is no stable fixed point and the system is chaotic
also for small $g$. This raises the interesting question about the
value of $\nu$ in biologically relevant networks. In biological application
of rate models, linear, quadratic or values of $\nu$ larger than
$1$ have been often used \cite{dayan2005theoretical,Troyer:1998ue,Heeger:1992jm,HanselSompBook1998,Albrecht:1995gy}.
The transfer function $\phi$ is often interpreted as reflecting the
\emph{f-I} curve of a spiking neuron. The linear leaky integrate and
fire (LLIF) model\textcolor{red}{{} }\cite{Tuckwell1988-Book,BrunelCourse}
exhibits a sharp ($1/|\log(x)$|) rise from zero, corresponding to
$\nu=0$. Our theory predicts that random networks with such\emph{
f-I} curves exhibit chaotic dynamics also for low $g$. In conductance
based (Hodgkin Huxley type) models the behavior of the \emph{f-I}
curve near threshold depends on the type and parameter values of the
various nonlinear conductances contributing to the spike generation.
In Hodgkin Huxley (HH) Type I models, often used for modeling cortical
neurons, the generic behavior of the \emph{f-I} curve near threshold
is a square root \textcolor{red}{\cite{RnzelErmentrout_HH}} making
them border-line cases for a transition at finite $g$. Similar $\nu=1/2$
behavior is exhibited also by nonlinear (NLLIF) models \cite{BrunelCourse}.
The presence of adaptation current with long adaptation time constant
results in a linearization of the \emph{f-I} curve \cite{ermentrout1998linearization,Shriki:2003cj},
and thus corresponds to $\nu=1$. Thus, slow adaptation currents in
randomly connected networks are predicted to stabilize fixed point
states at low $g$. It is interesting to note that the fixed point
equations for the population averaged activities (\ref{eq:mfM}) are
stable to linear perturbation even for $\nu<\frac{1}{2}$ due to the
smearing of the singularity near threshold by the gaussian fluctuations.
On the other hand, as we have shown here, this smearing is not strong
enough to avoid instability of the fixed point of the local activities
and fields.

Finally, our prediction that for $\nu\leq1/2$ no stable fixed point
exists was derived in the limit of large $K$ where the distribution
of local inputs is Gaussian, hence unbounded. However, in a finite
system, where $K$ is finite, there will be a finite gap between zero
and the smallest input (in absolute value). Hence, in a finite system
there should be a stable fixed point for sufficiently low values of
$g$, even for low values of $\nu$. Studying this finite size effect
is an interesting topic for future research. 

Although the properties of the transition to chaos are determined
by the behavior of $\phi(x)$ near threshold, the overall shape of
$\phi$ may also affect the system's behavior. For instance, in the
threshold linear case, the effect of changing the magnitude of the
external input is marginal, as it can be scaled out from the equations
determining the chaotic behavior, due to the inherent linearity above
threshold (see Eq (\ref{eq:SP_FP_sol})). In contrast, when $\phi(x)$
saturates to a finite value at large $x$, large external input pushes
neurons to saturation and suppresses the onset of chaos \cite{Rajan:2010tn}.
Also, the unboundedness of the threshold-linear $\phi$ leads to a
divergent dynamics for sufficiently large $g,$ see Fig. \ref{fig:Phase-diagram}.
Furthermore, when $\phi(x)$ grows exponentially with $x$ this instability
sets in as soon as the fixed point is unstable. This divergent dynamics
does not appear when $\phi$ has a finite saturation value.

\paragraph{Spiking dynamics}

The question whether networks with spiking dynamics exhibit a phase
transition to chaos at finite synaptic gain extends beyond the issue
of the shape of the \emph{f-I} curve. In contrast to recent claims
\cite{Ostojic:2014kd}, we have shown that a sharp transition from
a fixed point to chaos in such networks is meaningful only when synaptic
time constant is large, where there is a clear separation between
spiking dynamics and rate dynamics. In this limit, depending on the
shape of the $f-I$ curve, the underlying rates may be constant in
low synaptic gain and become chaotic above a critical gain, similar
to the behavior of rate based dynamics. The general correspondence
between smooth rate dynamic models and the dynamics of synaptic currents
in neuronal systems with long synaptic time constants, has been studied
previously \cite{ermentrout1994reduction}. However, the implications
on the transition to chaos in random neuronal networks were not previously
addressed. In any realistic systems the time constants are finite,
hence it is important to assess the smoothing of the transition due
to finite synaptic times. Here we have characterized this smoothing
by a scaling function with a single crossover exponent. This exponent
determines the rate of change from stochastic spiking dynamics with
static rates to smooth rate dynamics, as the synaptic time constant
increases. This scaling analysis predicts relatively strong 'finite
size' effects of the spiking dynamics near the transition of the corresponding
rate dynamics. First, the scaling regime is relatively large, given
by $|g-g_{c}|\propto(r\tau_{s})^{-2/5}$, where $r$ is the mean firing
rate and $\tau_{s}$ is the synaptic integration time. In addition,
the effective time constant of the synaptic fluctuations due to spiking,
scales as $\tau_{eff}\propto(r\tau_{s})^{1/5}$. Thus, a sharp transition
requires rather large values of $r\tau_{s}$. 

Concerning the biologically relevant regime, typical values of mean
rates range between an order of $1$Hz to $100$Hz. Fast synaptic
currents (AMPA and GABA\textsubscript{A}) have decay time constants
of the order of $1-10$ msec so are not expected to exhibit the above
transition. Slow synaptic currents (e.g., NMDA, GABA\textsubscript{B},
and peptide neurotransmitters) have time constants ranging from a
few hundred milliseconds to minutes \cite{Greengard:2001fv} and thus
might be appropriate candidates for exhibiting transition from spike
dynamics to rate fluctuations for systems with moderate rates (such
as primate visual and motor cortex).

Our analysis of networks of spiking neurons assumed inhomogeneous
Poisson spiking model, which is a well known and extensively used
statistical model of spiking fluctuations \cite{Brown:2002jy,Tuckwell1988-Book}.
However, it is interesting to ask whether our results hold for deterministic
spiking models with an appropriately smooth $f-I$ curve. We believe
our results regarding a transition to chaos for large $K$ and large
$r\tau_{s}$ are valid also for conductance based spiking dynamics,
since they rely only on the separation of time scales between the
firing dynamics and synaptic fluctuations. In particular, we expect
that for weak synaptic gain the network will be in an asynchronous
state characterize by synaptic inputs and population firing rates
whose fluctuation amplitude is small on the scale of $\tau_{s}$.
On the other hand, for strong synaptic gain, these fluctuations will
be large even on the scale of $\tau_{s}$ and the statistics of these
fluctuation will follow the chaotic dynamics of smooth rate dynamics. 

Beyond its simplicity, the advantage of the treatment the Poisson
model is that it demonstrates the condition for rate chaotic fluctuations
also for stochastically firing neurons. Our analysis of the Poisson
model was restricted to threshold-linear rate function. It is straightforward
to extend the DMFT equations to a general rate function, including
one with a power-law behavior above threshold with $\nu\neq1$ .It
will be of interest to study in more detail the effect of fast spiking
noise on the network dynamics, particularly for the cases where the
(noiseless) $f-I$ curve has $\nu<1/2.$ 

One difference between the Poisson and the deterministic spiking dynamics
is the degree of irregularity in the spike times, as measured e.g.,
by the standard deviation of the ISI distribution divided by its mean
(known as the coefficient of variation-CV). In the Poisson model,
the CV is close to $1$ in the 'fixed point' regime (as in homogeneous
Poisson process) and increases above it in the chaotic regime, due
to the induced rate fluctuations. In contrast, in deterministic spiking
models, the CV is expected to be substantially lower than $1$ in
the 'fixed point' regime. The detailed study of random networks with
deterministic spiking dynamics associated with sufficiently smooth
$f-I$ curve and slow synaptic time constants, is an interesting topic
for future studies. 

In this work, we have addressed the effect of spiking dynamics in
smearing the transition from a state with static underlying synaptic
currents to a state where the currents themselves (hence the underlying
rates) fluctuate in time. However, there can be other types of transitions
from non-chaotic dynamics to chaotic dynamics in a spiking network
\emph{even for fast synaptic time constants}. In the case of a deterministic
spiking networks, this transition may mark the separation from a limit
cycle at low $g$ to a chaotic attractor at high $g$. In this case,
chaos is typically of fast time constant and cannot be described as
instability in rates. In addition, it is likely that even in the non
chaotic regime of deterministic spiking models, irregular transients
(with effective negative Lyapunov exponent, termed as stable chaos)
will persist for a long time, and convergence time to the limit cycle
will grow exponentially with the system size (as in \cite{Zillmer:2009bm,Jahnke:2008fm,Zumdieck:2004be}).
Interestingly, the existence of chaos vs. stable chaos in spiking
networks has been shown to depend on the details of the spike initiation
dynamics \cite{monteforte2011single}. A transition from irregular
but non-chaotic state to a truly chaotic state might be important
for the information processing capacity of the system but will be
hard to observe experimentally, as it is not reflected in the properties
of the correlation functions.

Our discussion of biological relevance of rate based dynamics focused
on identifying the units in the rate based models as single neurons
and utilized temporal averaging as the mechanism for the adequacy
of a rate based theory. An alternative scenario where rate description
of a spiking network might be applicable is when the system consists
of clusters of neurons, where spatial averaging can lead to rate dynamics
(e.g. \cite{Shriki:2003cj,Ostojic:2011kf,Schaffer:2013hd}). Under
this interpretation, single units in our model represent weakly synchronous
neuronal subpopulations and the random connectivity corresponds to
the large scale effective connectivity between these populations.
As such these models can serve as a useful framework for investigating
aspects of the large scale nonlinear dynamics of the nervous system,
as measured by EEG and fMRI signals. 

This work explored the conditions under which random networks exhibit
a transition from fixed point to chaos and the rate of development
of chaotic fluctuations near the transition. These questions may bear
important consequences for the computations that such network can
produce. Recent models \cite{Sussillo2009544,Barak:2013bg,Mante:2013ie}
have shown that random recurrent networks can be shaped through learning
to generate a broad repertoire of desired trajectories with biologically
relevant time scales. These abilities prevail only near the onset
of chaos. For substantially low gain, the activity in the network
is strongly suppressed, whereas high above the transition, the intrinsic
fluctuations are too fast and erratic to match the smooth and slow
desired trajectories. It would be very interesting to study in detail
how the results of the present work affect the computational powers
of recurrent neuronal networks. 

\emph{Note added: }After the submission of this manuscript we became
aware of a manuscript \cite{Harish:2015je}that partially overlaps
with the present work.
\begin{acknowledgments}
We wish to thank L.F. Abbott and M. Stern for helpful discussions
and J.E. Fitzgerald and I. Landau for their comments on the manuscript.
We also wish to thank the anonymous reviewers for their the helpful
insights. This research was supported, in part, by the Gatsby Charitable
Foundation, the International Max Planck - Hebrew University Center,
the James S. McDonnell Foundation and the Simons Foundation (SCGB
{[}award 325207{]} to HS).
\end{acknowledgments}

\appendix

\section{DMFT equations for the autocorrelations\label{sec:app:DMFT-derivation}}

We extend previous derivations in \cite{Sompolinsky:1988ta} to the
current model of multiple subpopulation. Here, we show the derivation
of the self-consistent dynamical equations for the population averaged
autocorrelation function of the local fields, $\Delta_{k}(\tau)$,
defined in (\ref{eq:Autocorrelations}). The dynamic equations for
the fluctuations in the local fields, $\delta h_{k}^{i}(t)$ , are
given by (\ref{eq:mft}), where in the large $N$ limit, the noise
terms, $\eta_{k}^{i}(\tau)$, are Gaussian random variables with zero
mean and $\left\langle \eta_{k}^{i}(t)\eta_{k}^{i}(t+\tau)\right\rangle =\sum_{l}g_{kl}^{2}C_{l}(\tau)$,
where $C_{k}(\tau)\equiv\langle\phi(h_{k}^{i}(t))\phi(h_{k}^{i}(t+\tau))\rangle$
is calculated self-consistently by integrating over a temporally colored
Gaussian noise $\eta_{k}^{i}(\tau)$. A convenient way of treating
the temporal correlations between $\delta h_{k}(t)$ and $\delta h_{k}(t+\tau)$
is to introduce three uncorrelated Gaussian variables $y_{1}$, $y_{2}$
and $z$ with unit variances such that 
\begin{eqnarray*}
\delta h_{k}(t) & = & \sqrt{\alpha}y_{1}+\sqrt{\beta}z,\\
\delta h_{k}(t+\tau) & = & \sqrt{\alpha}y_{2}+\sqrt{\beta}z.
\end{eqnarray*}
with 
\begin{eqnarray*}
\alpha & = & \Delta_{k}(0)-\Delta_{k}(\tau),\\
\beta & = & \Delta_{k}(\tau).
\end{eqnarray*}
With these variables, the population averaged autocorrelation function
can be written as the integral over independent Gaussian variables
\begin{multline}
C_{k}(\tau)=\\
\intop Dy_{1}\intop Dy_{2}\intop Dz\phi\left(\sqrt{\alpha}y_{1}+\sqrt{\beta}z+u_{k}\right)\times\\
\phi\left(\sqrt{\alpha}y_{2}+\sqrt{\beta}z+u_{k}\right).\label{eq:app:C_independent_full}
\end{multline}

Finally, to yield the self-consistent equations for $\Delta_{k}$
it is convenient to use the Fourier transform of the dynamic equation
(\ref{eq:mft}) one gets 
\begin{equation}
(1+i\omega)\delta h_{k}^{i}(\omega)=\eta_{k}^{i}(\omega).\label{eq:app:DMFT_Fourier}
\end{equation}
from which we obtain, 
\begin{equation}
\left(1+\omega^{2}\right)\left\langle \left|\delta h_{k}^{i}(\omega)\right|^{2}\right\rangle =\left\langle \left|\eta_{k}^{i}(\omega)\right|^{2}\right\rangle =\sum_{l}g_{kl}^{2}C_{l}(\omega).
\end{equation}
Finally, preforming a Fourier transform back to the time domain, and
substituting the result of (\ref{eq:app:C_independent_full}), the
$P$ self consistent equations for the autocorrelations of the local
fields  read
\begin{multline*}
\left(1-\frac{\partial^{2}}{\partial\tau^{2}}\right)\Delta_{k}(\tau)=\\
\sum_{l}g_{kl}^{2}\left\langle \left\langle \phi\left(\sqrt{\Delta_{l}(0)-\Delta_{l}(\tau)}y+\sqrt{\Delta_{l}(\tau)}z+u_{l}\right)\right\rangle _{y}^{2}\right\rangle .
\end{multline*}

\section{Stability Equations for the Fixed Points\label{sec:App:UniformStability}}

In this Appendix we derive the stability conditions for the fixed
point states of the network.

\subsection{Population average linear response}

For simplicity we present the derivation with one population architecture.
The extension to multiple population is straightforward.

We evaluate the time dependent response function

\begin{equation}
\chi_{i}(t)\equiv\partial\phi(h_{i}(t))/\partial h^{0}(0)
\end{equation}

and denote by $\chi(t)$ the spatial average of $\chi_{i}$. 

at the fixed point solution. From the equations of motion, we obtain

\begin{equation}
(1+\partial_{t})\chi_{i}(t)=\phi'(h_{i})\sum_{j}\mathcal{J}_{ij}\chi_{j}(t)+\phi'(h_{i})\bar{g}\chi(t)+\phi'(h_{i})\delta(t)\label{eq:appUniform}
\end{equation}

where $\phi'(h_{i})$ are the derivatives of the activation function
evaluated at the fixed points. Let us consider the general term

\begin{equation}
\phi'(h_{i})\sum_{j}\mathcal{J}_{ij}\chi_{j}(t)=\frac{\partial}{\partial h(0)}\phi'(h_{i})\sum_{j}\mathcal{J}_{ij}\phi(h_{j}(t))
\end{equation}

Because the fixed point values of $h_{i}$ are independent of a perturbation
at time $0$ . Thus, using Eqs (\ref{eq:mft}) and (\ref{eq:mft_noise}),
we have 
\begin{equation}
\phi'(h_{i})\sum_{j}\mathcal{J}_{ij}\chi_{j}(t)=\frac{\partial}{\partial h(0)}(1+\partial_{t})\phi'(h_{i})\delta h_{i}(t)
\end{equation}

where $\partial_{t}$ stands for the time derivative operator. After
averaging and using mean field theory,

\begin{multline}
\langle\phi'(h{}_{i}(t_{2}))\delta h{}_{i}(t_{1})\rangle=\\
\langle\phi'(\sqrt{\Delta(t_{2},t_{2})-\Delta(t_{2},t_{1})}x+\sqrt{\Delta(t_{2},t_{1})}z+u(t_{2}))\times\\
(\sqrt{\Delta(t_{1},t_{1})-\Delta(t_{2},t_{1})}y+\sqrt{\Delta(t_{2},t_{1})}z)\rangle\\
=\langle\phi'(\sqrt{\Delta(t_{2},t_{2})-\Delta(t_{2},t_{1})}x+\\
\sqrt{\Delta(t_{2},t_{1})}z+u(t_{2}))\sqrt{\Delta(t_{2},t_{1})}z\rangle\\
=\Delta(t_{2,}t_{1})\langle\phi''(\sqrt{\Delta(t_{2},t_{2})}z+u(t_{2}))\rangle
\end{multline}

where $\Delta(t,t')=\langle\delta h(t)\delta h(t')\rangle$. Thus,
\begin{equation}
\langle\phi'\left(h_{i}\right)\sum_{j=1}^{N}\mathcal{J}_{ij}\chi_{j}(t)\rangle=(1+\partial_{t})\frac{\partial\Delta(t',t)}{\partial h^{0}(0)}\langle\phi''\rangle
\end{equation}

with $t'\rightarrow\infty$. Finally, we note that in this limit,
\begin{equation}
\frac{\partial\Delta(t_{2,}t)}{\partial h^{0}(0)}=\frac{\partial\langle\delta h(t_{2})\delta h(t)\rangle}{\partial h^{0}(0)}
\end{equation}

We denote 

\begin{equation}
\chi^{\Delta}(t)=2\chi^{\Delta}(\infty,t)
\end{equation}

where, 

\begin{equation}
\chi^{\Delta}(t',t)=\frac{\partial\Delta(t',t)}{\partial h^{0}(0)}
\end{equation}

Note that the factor of $2$ was introduced, because in the fixed
point, 

\begin{equation}
\chi^{\Delta}(t,t)=2\chi^{\Delta}(\infty,t)\equiv\chi^{\Delta}(t)\label{eq:DelDel}
\end{equation}
which is the susceptibility of the mean equal time variance. 

Substituting the above results in (\ref{eq:appUniform}) and averaging
we obtain,
\begin{equation}
(1-\langle\phi'\rangle\bar{g}+\partial_{t})\chi(t)=\frac{1}{2}\langle\phi''\rangle(1+\partial_{t})\chi^{\Delta}(t)+\langle\phi'\rangle\delta(t)\label{eq:sucseptibility1}
\end{equation}

\textbf{Calculating of $\chi^{\Delta}$}

From the DMFT we obtain

\begin{equation}
(1+\partial_{t_{1}})(1+\partial_{t_{2}})\Delta(t_{2,}t_{1})=g^{2}C(t_{1},t_{2})
\end{equation}

where,
\begin{multline}
C(t_{1},t_{2})=\langle\phi(\sqrt{\Delta(t_{2},t_{2})-\Delta(t_{2},t_{1})}x+\\
\sqrt{\Delta(t_{2},t_{1})}z+u(t_{2}))\phi(\sqrt{\Delta(t_{1},t_{1})-\Delta(t_{2},t_{1})}y+\\
\sqrt{\Delta(t_{2},t_{1})}z+u(t_{1}))\rangle\label{eq:appB:C_two_times}
\end{multline}

Thus,\begin{widetext} 
\begin{multline}
(1+\partial_{t_{1}})(1+\partial_{t_{2}})\chi^{\Delta}(t_{2,}t_{1})=g^{2}[\partial C(t_{1},t_{2})/\partial\Delta(t_{2},t_{1})\chi^{\Delta}(t_{2},t_{1})+\partial C(t_{1},t_{2})/\partial\Delta(t_{2},t_{2})\chi^{\Delta}(t_{2},t_{2})+\\
+\partial C(t_{1},t_{2})/\partial\Delta(t_{1},t_{1})\chi^{\Delta}(t_{1},t_{1})+\langle\phi'(t_{2})\phi(t_{1})\rangle[\bar{g}\chi(t_{2})+\delta(t_{2})]+\langle\phi(t_{2})\phi'(t_{1})\rangle[\bar{g}\chi(t_{1})+\delta(t_{1})]]
\end{multline}
\end{widetext}

where, by derivation of (\ref{eq:appB:C_two_times}),
\begin{equation}
\partial C(t_{1},t_{2})/\partial\Delta(t_{2},t_{1})=\langle\phi'(t_{2})\phi'(t_{1})\rangle
\end{equation}

\begin{equation}
\partial C(t_{1},t_{2})/\partial\Delta(t_{1},t_{1})=\frac{1}{2}\langle\phi(t_{1})\phi''(t_{1})\rangle
\end{equation}
and
\begin{equation}
\frac{\partial u}{\partial h^{0}}=\bar{g}\chi(t)+\delta(t)
\end{equation}

We will assume $t_{2}>t_{1}>t_{0}$ are all large but their difference
is of order 1.

Let us take $t_{2}-t_{1}\rightarrow\infty$ . Then, substituting (\ref{eq:DelDel})
one obtains: 

\begin{equation}
(1+\partial_{t})\chi^{\Delta}(t)=g^{2}[\langle\phi'^{2}\rangle+\langle\phi\phi''\rangle]\chi^{\Delta}(t)+2g^{2}\langle\phi\phi'\rangle[\bar{g}\chi(t)+\delta(t)]\label{eq:susceptibility2}
\end{equation}

Eqs (\ref{eq:sucseptibility1}) and (\ref{eq:susceptibility2}) constitutes
two coupled equations for the response functions of the mean and variance
of the population activity. It is perhaps convenient to eliminate
the time derivative in the RHS of (\ref{eq:sucseptibility1}) by substituting
in it Eq (\ref{eq:susceptibility2}), resulting in 

\begin{multline}
(1-\langle\phi'\rangle\bar{g}+\partial_{t})\chi(t)=\frac{1}{2}\langle\phi''\rangle g^{2}[\langle\phi'^{2}\rangle+\langle\phi\phi''\rangle]\chi^{\Delta}(t)+\\
\langle\phi''\rangle g^{2}\langle\phi\phi'\rangle[\bar{g}\chi(t)+\delta(t)]+\langle\phi'\rangle\delta(t)\label{eq:susceptibility3}
\end{multline}

We can write these equations by

\begin{equation}
(a+\partial_{t})\chi(t)=b\chi(t)+c\delta(t)
\end{equation}
\begin{equation}
(d+\partial_{t})\chi^{\Delta}(t)=e\chi(t)+f\delta(t)
\end{equation}

or, in Fourier and matrix representation, 

\begin{equation}
\begin{bmatrix}(1-\bar{g}a+i\omega) & -b\\
-\bar{g}e & (1-d+i\omega)
\end{bmatrix}\begin{bmatrix}\chi\\
\chi^{\Delta}
\end{bmatrix}=\begin{bmatrix}a\\
e
\end{bmatrix},
\end{equation}

where

\begin{equation}
a=\langle\phi''\rangle g^{2}\langle\phi\phi'\rangle+\langle\phi'\rangle,
\end{equation}

\begin{equation}
d=g^{2}[\langle\phi'^{2}\rangle+\langle\phi\phi''\rangle],
\end{equation}
\begin{equation}
b=\frac{1}{2}\langle\phi''\rangle g^{2}[\langle\phi'^{2}\rangle+\langle\phi\phi''\rangle],
\end{equation}
and
\begin{equation}
e=2g^{2}\langle\phi\phi'\rangle\bar{g}.
\end{equation}

\paragraph{Multiple population network}

A straightforward generalization to the multiple population case yields
for 

\begin{equation}
\chi_{kl}(t)\equiv\frac{\partial m_{k}(t)}{\partial h_{l}(0)}
\end{equation}

and

\begin{equation}
\chi_{kl}^{\Delta}(t)=\frac{\partial}{\partial h_{l}(0)}\Delta_{k}(t).
\end{equation}
In this case, the equation for $\chi^{\Delta}(t)$( Eq (\ref{eq:susceptibility2}))
reads
\begin{multline}
(1+\partial_{t})\chi_{kl}^{\Delta}(t)=\\
\sum_{m}g_{km}^{2}\left[\langle\phi_{m}'^{2}\rangle+\langle\phi_{m}\phi_{m}''\rangle\right]\chi_{ml}^{\Delta}(t)\\
+2\sum_{m}g_{km}^{2}\langle\phi_{m}\phi_{m}'\rangle\left[\sum_{m'}\bar{g}_{mm'}\chi_{m'l}(t)+\delta_{ml}\delta(t)\right]
\end{multline}
or, in matrix notation,
\begin{equation}
(\mathbf{I}-\mathbf{D}+\partial_{t})\chi^{\Delta}=\mathbf{E}\left(\bar{\mathbf{g}}\chi+\mathbf{I}\delta(t)\right).\label{eq:AppB:chiDeltaMatrix}
\end{equation}
Likewise, the dynamical equation on $\chi(t)$ is 
\begin{multline}
(1+\partial_{t})\chi_{kl}(t)=\\
\sum_{m}\bar{g}_{km}\langle\phi'_{m}\rangle\chi_{ml}(t)+\langle\phi_{k}'\rangle\delta_{kl}\delta(t)\\
+\frac{1}{2}\langle\phi_{k}''\rangle(1+\partial_{t})\chi_{kl}^{\Delta}(t)
\end{multline}
or
\begin{equation}
(\mathbf{I}-\mathbf{A}\bar{\mathbf{g}}+\partial_{t})\chi(t)=\mathbf{B}\chi^{\Delta}+\mathbf{A}\delta(t).\label{eq:AppB:chiMatrix}
\end{equation}
Where in the above we have defined the $P\times P$ matrices 
\begin{equation}
A_{kl}=\delta_{kl}\langle\phi'_{l}\rangle+\langle\phi_{k}''\rangle g_{kl}^{2}\langle\phi_{l}\phi_{l}'\rangle,
\end{equation}
\begin{equation}
B_{kl}=\frac{1}{2}g_{kl}^{2}\langle\phi_{k}''\rangle\left[\langle\phi_{l}'^{2}\rangle+\langle\phi_{l}\phi_{l}''\rangle\right],
\end{equation}
\begin{equation}
D_{kl}=g_{kl}^{2}\left[\langle\phi_{l}'^{2}\rangle+\langle\phi_{l}\phi_{l}''\rangle\right],
\end{equation}
and
\begin{equation}
E_{kl}\equiv2g_{kl}^{2}\langle\phi_{l}\phi_{l}'\rangle.
\end{equation}
Eq (\ref{eq:AppB:chiDeltaMatrix}) and (\ref{eq:AppB:chiMatrix})
can be written using a $2P\times2P$ matrix in Fourier space as

\begin{equation}
\begin{bmatrix}(\mathbf{I}-\mathbf{A}\bar{\mathbf{g}}+i\omega) & -\mathbf{B}\\
-\mathbf{E}\mathbf{\bar{g}} & (\mathbf{I}-\mathbf{D}+i\omega)
\end{bmatrix}\begin{bmatrix}\chi(\omega)\\
\chi^{\Delta}(\omega)
\end{bmatrix}=\begin{bmatrix}\mathbf{A}\\
\mathbf{E}
\end{bmatrix}\label{eq:uniformstabeq-1}
\end{equation}

Thus, the fixed point is stable against population average perturbations
provided that all the eigenvalues of the matrix 
\begin{equation}
\begin{bmatrix}\mathbf{I}-\mathbf{A}\bar{\mathbf{g}} & -\mathbf{B}\\
-\mathbf{E}\bar{\mathbf{g}} & 1-\mathbf{D}
\end{bmatrix}\label{eq:uniformStabDet-1}
\end{equation}

have negative real part.

\subsection{Stability of fixed point to local perturbations }

The susceptibility matrix of the network to a random local perturbation
is defied as 

\begin{equation}
\chi_{kl}^{ij}(t-t')=\partial h_{k}^{i}(t)/\partial h_{l}^{0j}(t')
\end{equation}
where $h_{l}^{0j}(t')$ are infinitesimal external perturbations around
the fixed point state. We are interested in the mean square of this
quantity. Using the dynamical equation (\ref{eq:Full_model}), we
find the time evolution of a the susceptibility\begin{widetext} 
\begin{multline}
\left(1+\frac{d}{dt}\right)\chi_{kl}^{ij}(t-t')=\\
\sum_{l'=1}^{P}\sum_{j'=1}^{N}\mathcal{J}_{kl'}^{ij'}\phi'\left(h_{l'}^{j'}\right)\chi_{l'l}^{j'j}(t-t')+\frac{1}{N}\sum_{l'=1}^{P}\bar{g}_{kl'}\sum_{j'}\phi'\left(h_{l'}^{j'}\right)\chi_{l'l}^{j'j}(t-t')+\delta(t-t')\delta_{ij}\delta_{kl},\:t\geq0\label{eq:app:chi_dynamics-1}
\end{multline}
\end{widetext} where $\delta_{ij}$ and $\delta(t)$ are the Kronecker
and Dirac delta functions respectively. Defining, 

\begin{equation}
G_{kl}(\omega_{1,},\omega_{2})=N^{-1}\sum_{ij}\langle\chi_{kl}^{ij}(\omega_{1})\chi_{kl}^{ij}(\omega_{2})\rangle\label{eq:localG-1}
\end{equation}

and averaging (\ref{eq:app:chi_dynamics-1}) yields

\begin{equation}
\mathbf{G}(\omega_{1,},\omega_{2})=[(1+i\omega_{1})(1+i\omega_{2})I-\mathbf{M}]^{-1}\label{eq:stabFP-1}
\end{equation}

where $\mathbf{M}$ is the \emph{stability matrix,} $M_{kl}=\langle\phi'(h_{k}^{i})\phi'(h_{l}^{i})\rangle$
(Eq (\ref{eq:stability_matrix_M})). 

Note that $G_{kl}$ are of order 1; the contribution of the uniform
$\chi_{kl}$ to them is of order $1/N$ hence negligible. Furthermore,
the uniform components of the interactions (in Eq. (\ref{eq:app:chi_dynamics-1}))
do not contribute to (\ref{eq:localG-1}) to leading order as they
are smaller by a factor of $1/N$ relative to the random contributions.

\section{Stability Equations and the largest Lyapunov Exponent (LE)\label{sec:Appendix_Stability}}

\global\long\def\bra#1{\left\langle #1\right|}
\global\long\def\ket#1{\left|#1\right\rangle }

The susceptibility matrix of the network to a random local perturbation
is defied as 

\begin{equation}
\chi_{kl}^{ij}(t,t')=\partial h_{k}^{i}(t)/\partial h_{l}^{0j}(t')
\end{equation}
where $h_{l}^{0j}(t')$ are infinitesimal external perturbations.
Using the dynamical equation (\ref{eq:Full_model}), we find the time
evolution of a the susceptibility 
\begin{multline}
\left(1+\frac{d}{dt}\right)\chi_{kl}^{ij}(t,t')=\\
\sum_{l'=1}^{P}\sum_{j'=1}^{N}\mathcal{J}_{kl'}^{ij'}\phi'\left(h_{l'}^{j'}(t)\right)\chi_{l'l}^{j'j}(t,t')+\\
\frac{1}{N}\sum_{l'=1}^{P}\bar{g}_{kl'}\sum_{j'}\phi'\left(h_{l'}^{j'}(t)\right)\chi_{l'l}^{j'j}(t,t')+\\
\delta(t-t')\delta_{ij}\delta_{kl},\:t\geq0\label{eq:app:chi_dynamics}
\end{multline}
where $\delta_{ij}$ and $\delta(t)$ are the Kronecker and Dirac
delta functions respectively, denoting a local spatiotemporal perturbation.

\paragraph{\textbf{Population averaged susceptibility}}

Let us define$\chi^{0}$ as the within-population spatial average
susceptibility

\begin{equation}
\chi_{kl}^{0}(t-t')=\frac{1}{N^{2}}\sum_{i,j=1}^{N}\chi_{kl}^{ij}(t,t')=\frac{1}{N^{2}}\sum_{i,j=1}^{N}\langle\chi_{kl}^{ij}(t,t')\rangle_{\mathcal{J}}\label{eq:UniformChi}
\end{equation}

This uniform matrix is formally given by

\begin{equation}
\chi^{0}(t-t')=\langle\left[I-\tilde{\mathcal{J}}-\frac{1}{N}\tilde{J}^{0}\right]^{-1}\rangle a(t)\delta(t-t'),\;t\geq t'
\end{equation}

where $I$ is spatial identitiy matris, $a(t)$ is the temporal operator
$a=(1+d/dt)^{-1}$, $\tilde{\mathcal{J}_{kl}^{ij}}=a\mathcal{J}_{kl}^{ij}\phi'\left(h_{l'}^{j'}\right)$
and $\tilde{J}{}_{kl}^{0ij}=a\bar{g}_{kl}\phi'\left(h_{l}^{j}\right)$
. {[}The RHS is formally an $NPxNP$ matrix; however, it is meant
to be reduced to a $PxP$ matrix by averaging over the $i,j$ indices{]}. 

Expanding the RHS in powers of $\tilde{\mathcal{J}}$ shows that all
contributions vanish upon averaging, hence

\begin{equation}
\chi^{0}(t-t')=\langle\left[I-\frac{1}{N}\tilde{J}^{0}\right]^{-1}\rangle a(t)\delta(t-t'),\;t\geq t'\label{eq:averageUniformChi}
\end{equation}

Furthermore, performing the averaging in the RHS of \ref{eq:averageUniformChi}
and fourier transforming the result yields 
\begin{equation}
\chi^{0}(\omega)=\frac{1}{N}[(1+i\omega)I-J^{0}]^{-1}
\end{equation}

where
\begin{equation}
J_{kl}^{0}=\bar{g}_{kl}\langle\phi'(h_{l}^{i})\rangle
\end{equation}

Note that this population average susceptibility is the solution of
the population averaged part of the dynamics, namely,

\textcolor{black}{
\begin{equation}
\left(1+\frac{d}{dt}\right)\chi_{kl}^{0}(t)=\sum_{k'}\bar{g}_{kk'}\left\langle \phi'\left(h_{i}^{l}\right)\right\rangle \chi_{k'l}^{0}(t)+\frac{1}{N}\delta_{kl}\delta(t),\,t\geq0\label{eq:app_Stab_Uniform-1}
\end{equation}
}

Note that all elements of $\chi_{kl}^{0}$ are of order $1/N$ including
the diagonal. This is becuase $\chi_{kk}^{0ii}$which is of order
$1$ adds only a $1/N$ contribution to the spatially averaged quantity,
\ref{eq:UniformChi}.

\paragraph{\textbf{Statistics of the full susceptibility}}

We now turn to evaluate the full susceptibility,$\chi_{kl}^{ij}$.
The random fluctuations in the susceptibility, i.e, the off diagnoal
elements of $\chi$ are of order $1/\sqrt{N}$ as will be seen below,
i.e., locally they are much larger than$\chi_{kl}^{0}$. Their average
second order moments are defined in terms of 
\begin{equation}
G_{kl}\left(t_{a},t_{b},t_{c},t_{d}\right)=\frac{1}{N}\sum_{i,j}^{N}\langle\chi_{kl}^{ij}(t_{a},t_{c})\chi_{kl}^{ij}(t_{b},t_{d}\rangle_{\mathcal{J}},\label{eq:Goperator}
\end{equation}
To evaluate \ref{eq:Goperator}, we multiply two realizations of Eq
(\ref{eq:app:chi_dynamics}) (differing in the time indices) and take
the spatial average over all neurons, leading to \begin{widetext}
\begin{multline}
\left(1+\frac{\partial}{\partial t_{a}}\right)\left(1+\frac{\partial}{\partial t_{b}}\right)G_{kl}\left(t_{a},t_{b},t_{c},t_{d}\right)-\sum_{m}M_{kl}(t_{a}-t_{b})G_{ml}\left(t_{a},t_{b},t_{c},t_{d}\right)=\\
\delta\left(t_{a}-t_{b}-t_{c}+t_{d}\right)\delta\left(t_{a}+t_{b}-t_{c}-t_{d}\right)\delta_{kl}.\label{eq:app:G_abcd_dynamics}
\end{multline}
\end{widetext} where

\begin{equation}
M_{kl}(\tau)=g_{kl}^{2}\left\langle \phi'(h_{i}^{l}(t+\tau)\phi'(h_{i}^{l}(t)\right\rangle .
\end{equation}

Note that $G_{kl}$ are of order ; the contribution of $\chi_{kl}^{0}$
to them is of order $1/N$ hence negligible. Furthermore, the uniform
components of the interactions (in Eq. \ref{eq:app:chi_dynamics})
do not contribute to \ref{eq:Goperator} to leading order as they
are smaller by a factor of $1/N$ relative to the random contributions. 

Defining new time variables as $\tau=t_{a}-t_{b}$, $\tau'=t_{c}-t_{d}$,
$T=t_{a}+t_{b}$ and $T'=t_{c}+t_{d}$, equation (\ref{eq:app:G_abcd_dynamics})
can be written as 
\begin{equation}
\left[\left(\left(1+\frac{\partial}{\partial T}\right)^{2}-1\right)I+\mathcal{H}(\tau)\right]G=2\delta(T-T')\delta(\tau-\tau')\mathbf{I},\label{eq:app:G_dynamics}
\end{equation}
 where $\mathcal{H}(\tau)$ is the $P\times P$ operator,

\begin{equation}
\mathcal{H}(\tau)=-\frac{\partial^{2}}{\partial\tau^{2}}\mathbf{I}+\mathbf{I}-\mathbf{M}(\tau),
\end{equation}
and 

(see (\ref{eq:Hamiltonian})).

\paragraph{Fixed point}

At the fixed point, 

\begin{equation}
\mathcal{H}(\tau)=-\frac{\partial^{2}}{\partial\tau^{2}}\mathbf{I}+\mathbf{I}-\mathbf{M}
\end{equation}
where $M$ is time independent $PxP$ matrix. Fourier transforming
\ref{eq:app:G_dynamics} yields,

\begin{equation}
\left[(2i\text{\textgreek{W}}-\text{\textgreek{W}}^{2})I+(\omega^{2}+1)I-M\right]G(\text{\textgreek{W}},\omega)=4\pi\mathbf{I},\label{eq:app:G_dynamics-fp}
\end{equation}

where $\text{\textgreek{W}}$ and $\omega$ are the frequencies associated
with $T$ and $\tau$ respectively,

\begin{equation}
G(\text{\textgreek{W}},\omega)=\int dT\int d\tau G(T,0,\tau,0)\exp(-i\text{\textgreek{W}}T-i\omega\tau)
\end{equation}
. Thus, the zero frequency limit, yields a static matrix given (up
to a constant) by 
\[
\mathbf{G(}\text{\textgreek{W}=0},\omega=0)=(\mathbf{I}-\mathbf{M})^{-1},
\]
where $\mathbf{M}$ is defined in (\ref{eq:M_kl-def}) and fixed point
stability requires that all eigenvalues of $M$ have real part less
than one. Note that $\mathbf{G}_{kl}=N\left\langle (\chi_{kl}^{ij}(\omega=0))^{2}\right\rangle $,
i.e., the square average of the local static susceptibility.

\paragraph{Chaotic regime}

The largest Lyapunov exponent is the exponential divergent rate of
the squared susceptibility$N\left\langle (\chi_{kl}^{ij}(t,0))^{2}\right\rangle =G_{kl}(2t,0,0,0)$,
in the chaotic state, which corresponds to $T=2t,$ and $T'=\tau=\tau'=0$.
Thus, the largest Lyapunov exponent is found by

\begin{equation}
\lambda_{L}=\lim_{t\to\infty}\frac{1}{2t}\ln\sum_{kl}^{P}G_{kl}(2t,0,0,0).
\end{equation}

To evaluate it, we first write the time dependent solution of (\ref{eq:app:G_dynamics}),
as 
\begin{equation}
G_{kl}(T,T',\tau,\tau')=\sum_{\mu=1}^{\infty}\Gamma_{\mu}(T,T')\ket{\psi_{k}^{\mu}(\tau)}\bra{\psi_{l}^{\mu}(\tau')},
\end{equation}
where $\bra{\psi^{\mu}(\tau)}$ and $\ket{\psi^{\mu}(\tau})$ are
the left and right $\mu-th$ eigenvectors of the Hamiltonian $\mathcal{H}(\tau)$.
We note that in general $\mathcal{H}(\tau)$ is non-hermitic and the
eigenvectors $\ket{\psi^{\mu}(\tau)}$ are not necessarily orthogonal.
From (\ref{eq:app:G_dynamics}), we find that the general solution
for $\Gamma_{\mu}(T,T')$ is proportional to $e^{\lambda_{\mu}(T-T')}$
where 
\[
\lambda_{\mu}=-1\pm\sqrt{1-\epsilon_{\mu}}
\]
and $\epsilon_{\mu}$is the corresponding eigenvalue of $\mathcal{H}(\tau)$.
For completeness, one must consider the boundary conditions on the
solutions, namely that solution exists only for $T>T'$ and that the
second derivative of $\mathbf{G}$ is a delta function while the first
derivative is finite. One obtains
\begin{multline}
G_{kl}(T,T',\tau,\tau')=\\
4\sum_{\mu=1}^{\infty}\frac{\Theta\left(T-T'\right)\ket{\psi_{k}^{\mu}(\tau)}\bra{\psi_{l}^{\mu}(\tau')}}{\sqrt{1-\epsilon_{\mu}}}e^{-(T-T')}\\
\times\sinh\left((T-T')\sqrt{1-\epsilon_{\mu}}\right)\label{eq:Gamma_sol}
\end{multline}
Finally,
\begin{multline}
N\langle\chi_{kl}^{2}(t,0)\rangle=G_{kl}(2t,0,0,0)=\\
4\sum_{\mu=1}^{\infty}\frac{\ket{\psi_{k}^{\mu}(0)}\bra{\psi_{l}^{\mu}(0)}}{\sqrt{1-\epsilon_{\mu}}}e^{-2t}\sinh\left(2t\sqrt{1-\epsilon_{\mu}}\right).
\end{multline}
Thus, the maximal Lyapunov exponent is given by 
\begin{equation}
\lambda_{L}=-1+\sqrt{1-\epsilon_{0}},
\end{equation}
where $\epsilon_{0}$ is the ground state of the Hamiltonian (\ref{eq:Hamiltonian}).

\section{Exponential transfer function\label{sec:Exponential-transfer-function}}

We show that a single inhibitory population, with the architecture
studied in Section \ref{sec:DMFT} and an exponential transfer function,
$\phi(x)=e^{x}$does not exhibit a chaotic phase. The fixed point
of the dynamic is stable, when $g^{2}C'(x)<1$, where $C=\left\langle \phi_{i}^{2}\right\rangle $,
leading to 

\begin{equation}
g^{2}\left\langle \exp\left(\sqrt{\Delta_{0}}z+u\right)\right\rangle =g^{2}e^{2u+\Delta_{0}}=g^{2}m^{2}<1.\label{eq:app:exp_stability}
\end{equation}
The variance, given by $\Delta_{0}=g^{2}C(x)$, is 
\begin{equation}
\Delta_{0}=g^{2}\left\langle exp\left(2\sqrt{\Delta_{0}z}+2u\right)\right\rangle =g^{2}e^{2u+\Delta_{0}}=g^{2}m^{2}.\label{eq:app:exp_FP}
\end{equation}

It follows from (\ref{eq:app:exp_stability}) and (\ref{eq:app:exp_FP})
that the fixed point loses its stability, at the critical point $g_{c}$
where $\Delta_{0}=g^{2}m^{2}=1$. For variance values higher than
$1$, the dynamics is characterized by the autocorrelation function
given by the solution the dynamical MF differential equation in (\ref{eq:one_pop_EOM}).
A stable chaotic solution requires 
\begin{equation}
\left.\frac{\partial^{2}}{\partial\tau^{2}}\Delta\right|_{\tau=\infty}=0,
\end{equation}
or, equivalently, the vanishing of the potential (\ref{eq:one_pop_potential})
at $\tau=\infty$. Carrying the gaussian integrals in (\ref{eq:one_pop_potential})
one finds, that if the potential vanish then 
\begin{equation}
\Delta_{0}=\Delta(\infty)e^{3\Delta(\infty)}.\label{eq:exp_pot}
\end{equation}
Since every solution for the autocorrelation function require $0\leq\Delta(\infty)\leq\Delta_{0}$,
Eq (\ref{eq:exp_pot}), can not be obeyed for a non vanishing variance,
and no chaotic phase exist. Consequentially, when $\Delta_{0}>1$,
the mean activity diverges.

\section{Dynamic equation near criticality\label{sec:Appendix_Rescaling}}

\paragraph{Single population}

Using the homogeneity of the transfer function, the correlation function
$C(\tau)$ (\ref{eq:dMFT_C}) can be written in terms of the parameter
$\tdel(\tau)=1-\Delta(\tau)/\Delta_{0}$ giving 
\begin{equation}
C(\tau)=\Delta_{0}^{2\nu}\left\langle \left\langle \phi\left(\sqrt{\tdel(\tau)}y+\sqrt{1-\tdel(\tau)}z+x\right)\right\rangle _{y}^{2}\right\rangle _{z}.
\end{equation}
Near and above criticality $C(\tau)$ can be expanded in powers of
$\tdel(\tau)$ about the fixed point value $\tdel(\tau)=0$. Since
linear analysis is not sufficient to account for the critical phenomena
we keep all terms up to the first non linear term in $\tdel(\tau)$,
\begin{multline}
\Delta_{0}^{-2\nu}C=\left\langle \phi_{\nu}^{2}(z+x)\right\rangle -\nu^{2}\left\langle \phi_{\nu-1}^{2}(z+x)\right\rangle \tdel+c\tdel^{\rho},\label{eq:C_critical_expan}
\end{multline}
 where $c$ is a numerical factor of order $1$. The gaussian integral
in $\left\langle \phi_{\nu-n}^{2}\right\rangle $ diverges for $2\left(\nu-n\right)\leq-1$.
As a result, the first nonlinear term may be larger than quadratic.
The first non linear term is $\rho=2$ for $\nu>\frac{3}{2}$ and
$\rho=\frac{3}{2}$ for $\frac{1}{2}<\nu\leq\frac{3}{2}$.

For convenience, we redefine the gain parameter as $\tilde{g}=g\Delta_{0}^{\left(\nu-1\right)/2}$;
the value of the gain $g$ can be recovered using the mean field equations,
(\ref{eq:m}). The dynamic equation (\ref{eq:one_pop_EOM})
can be written as
\begin{multline}
\frac{\partial^{2}}{\partial\tau^{2}}\tdel=\tilde{g}^{2}\left\langle \phi_{\nu}^{2}(z+x)\right\rangle -1+\\
\left(1-\tilde{g}^{2}\nu^{2}\left\langle \phi_{\nu-1}^{2}(z+x)\right\rangle \right)\tdel+\tilde{g}^{2}c\tdel^{\rho}.\label{eq:app:delta_dynamics}
\end{multline}
Next, we expand each term in (\ref{eq:app:delta_dynamics}) in powers
of $\epsilon=g^{2}/g_{c}^{2}-1$, which controls the distance from
criticality. We also denote $\delta x=x-x_{c}$ which is a function
of $\epsilon$. Eq (\ref{eq:app:delta_dynamics}) can be written as
\begin{equation}
\frac{\partial^{2}\tdel(\tau)}{\partial\tau^{2}}=a\left(\epsilon\right)+b(\epsilon)\tdel(\tau)+c\tdel^{\rho}(\tau),\label{eq:app:delta_dynamics_reduced}
\end{equation}
where
\begin{multline}
a(\epsilon)=\epsilon+2\nu\tilde{g}_{c}^{2}\left\langle \left[z+x_{c}\right]_{+}^{2\nu-1}\right\rangle \delta x(1+\epsilon)+\\
2\nu\left(2\nu-1\right)\tilde{g}_{c}^{2}\left\langle \left[z+x_{c}\right]_{+}^{2\nu-2}\right\rangle \delta x^{2}+\mathcal{O}\left(\epsilon\delta x^{2}\right)\label{eq:app:a}
\end{multline}
and
\begin{equation}
b(\epsilon)=\epsilon+2\tilde{g}^{2}\nu^{2}b_{\nu}^{(1)}\delta x.\label{eq:app:b}
\end{equation}

The factor $b_{\nu}^{(1)}$ is of order $1$ and equals
\begin{equation}
b_{\nu}^{(1)}=\begin{cases}
(\nu-1)\left\langle \left[z+x_{c}\right]_{+}^{2\nu-3}\right\rangle  & \nu>1\\
\frac{1}{\sqrt{2\pi}} & \nu=1\\
\left(\left\langle \left[z+x_{c}\right]_{+}^{2\nu-1}\right\rangle +\left\langle \left[z+x_{c}\right]_{+}^{2\nu-2}\right\rangle \right) & \frac{1}{2}<\nu<1
\end{cases}.\label{eq:app:b-1}
\end{equation}

In Eq (\ref{eq:app:a}) through (\ref{eq:app:b-1}) we have used the
stability condition (\ref{eq:stabFP}) 
\begin{equation}
\nu^{2}\tilde{g}_{c}^{2}\left\langle \phi_{\nu-1}^{2}\left(z+x_{c}\right)\right\rangle =1,
\end{equation}
and the fixed point solution for the variance, which also holds at
the transition point, (\ref{eq:FP_delta0})
\begin{equation}
\tilde{g}_{c}^{2}\left\langle \phi_{\nu}^{2}\left(z+x_{c}\right)\right\rangle =1.
\end{equation}

We expect that $a(\epsilon)$ would scale as $b(\epsilon)\tdel,$
implying that the leading order in (\ref{eq:app:a}) vanish and
\begin{equation}
\delta x=-\frac{1}{2\nu}\tilde{g}_{c}^{-2}\left\langle \left[z+x_{c}\right]_{+}^{2\nu-1}\right\rangle ^{-1}\epsilon.\label{eq:appLDeltax}
\end{equation}
 Thus, we can write $b(\epsilon)=\epsilon b$. Following the same
argument as in (\ref{eq:one_pop_potential}), we define a classical
potential for $\tdel(\tau)$  by integrating the LHS of (\ref{eq:app:delta_dynamics_reduced}),
giving 
\begin{equation}
V\left[\tdel\right]=a\left(\epsilon\right)\tdel+\frac{1}{2}\epsilon b\tdel^{2}+\frac{c}{\rho+1}\tdel^{\rho+1}.
\end{equation}
By applying the boundary requirements for a stable chaotic solution,
$V[0]=V\left[\tdel(\infty)\right]$ and $\left.\frac{dV}{d\tdel}\right|_{\tau=\infty}=0$
one gets the scaling of $\tdel$ and $a$ for small $\epsilon$ 
\begin{equation}
\tdel(\infty)=\left(\frac{(1+\rho)}{2\rho}\frac{b}{c}\epsilon\right)^{\frac{1}{\rho-1}}
\end{equation}
and
\begin{equation}
a=-\frac{\rho-1}{\rho+1}c\tdel^{\rho}(\infty)\sim\epsilon^{\frac{\rho}{\rho-1}}.\label{eq:app:a_scale}
\end{equation}

\paragraph{Multiple populations}

In a network composed of $P$ populations, stability is determined
by the eigenvalues of the $PxP$ stability matrix, (\ref{eq:Hamiltonian}).
Here we define a normalized stability matrix $\tilde{M}_{kl}=\Delta_{k}^{-1}(0)M_{kl}=\tilde{g}_{kl}^{2}\left\langle \phi'^{2}(z+x_{k}\right\rangle $
and 
\begin{equation}
\tilde{g}_{kl}^{2}=g_{kl}^{2}\frac{\Delta_{l}(0)}{\Delta_{k}(0)}.
\end{equation}
We write $\tilde{M}_{kl}=\sum_{\mu}\Lambda_{\mu}R_{k}L_{l}^{*}$ where
$\Lambda_{\mu}$are the eigenvalues ordered according to decreasing
value of their real part, and $R$,$L$ are the right and left eigenvectors
of $\mathbf{\tilde{M}}$ respectively.  At the transition, $\Lambda_{1}=1$
and $Re\Lambda_{\mu\neq1}\leq1$. When $g_{kl}\approx g_{kl}^{c}$,
the unstable eigenvalue is $\Lambda_{1}=1+\epsilon$ where $|\epsilon|\ll1$
and the sign of $\epsilon$ determines which direction in the space
of $g$ (away from $g_{kl}^{c}$) is the unstable one. To leading
order, the eigenvector $R_{1}$does not change. Note that $\mathbf{\Lambda_{1}}$
depends on both $g$ and $x$. For convenience, we define $\epsilon$
via $\Lambda_{1}(g,x_{c}$) i.e., as the change induced in $\Lambda_{1}$due
to change in $g$ for $x=x_{c}$. We are interested in the properties
of the system for $0<\epsilon\ll1$. For multiple sub-population,
equation (\ref{eq:app:delta_dynamics_reduced}) takes the form

\begin{multline}
\frac{\partial^{2}\tdel_{k}}{\partial\tau^{2}}=a_{k}(\epsilon)+\sum_{kl}(\mathbf{I}-\tilde{\mathbf{M}})_{kl}\tdel_{l}+\sum\tilde{g}_{kl}^{2}c_{l}\tdel_{l}^{\rho},\label{eq:app:delta_dynamics_multi}
\end{multline}
where $a_{k}=1-\sum_{l}\tilde{g}_{kl}^{2}C\left\langle \phi^{2}(z+x_{k}\right\rangle $,
which vanishes at criticality due to (\ref{eq:FP_delta0}), and $\tilde{M}_{kl}=\tilde{g}_{kl}^{2}\left\langle \phi'^{2}(z+x_{k}\right\rangle $.
Since the matrix $\mathbf{I}-\tilde{\mathbf{M}}$ vanishes only in
the $1-$ direction, the dominant component of the vector $\tdel$
is in this direction. To see this, we make the ansatz $\tdel_{k}(\tau)\equiv1-\Delta_{k}(\tau)/\Delta_{k}^{0}=\sum_{\mu}\zeta^{\mu}(\tau)R_{\mu}^{k}$
and assume that $|\zeta^{1}|\gg|\zeta^{\mu}|$. We then obtain, 

\begin{eqnarray}
\frac{\partial^{2}\zeta^{1}}{\partial\tau^{2}} & = & \hat{a}^{1}(\epsilon)+(1-\Lambda_{1})\zeta^{1}(\tau)+\hat{c}^{1}\zeta^{1}(\tau)^{\rho},\\
\frac{\partial^{2}\zeta^{\mu}}{\partial\tau^{2}} & = & \hat{a}^{\mu}(\epsilon)+(1-\Lambda_{\mu})\zeta^{1}(\tau)+\hat{c}^{\mu}\zeta^{1}(\tau)^{\rho}\;\mu\ne1,
\end{eqnarray}
i

where $\hat{a}$ and $\hat{c}$ are the factors in the basis of $\mathbf{\tilde{M}}$.
Here $1-\Lambda_{1}$vanishes at the transition and is of the order
of $\epsilon,\delta x$, while $(1-\Lambda_{\mu})$ remains of order
$1$ near the transition. Thus, $\zeta^{\mu}(\tau)$ scales as $\zeta^{1}(\tau)^{\rho}\ll\zeta(\tau)$,
and the ``normal form'' of the dynamics of $\tdel_{k}$, Eq (\ref{eq:app:delta_dynamics_multi}),
is the same as that of a single population, Eq (\ref{eq:app:delta_dynamics_reduced}).
In addition, in the generic case, we expect all $\tdel_{k}$ to have
nonzero projection on $R_{1}$. Hence, they will all scale as $\zeta^{1}$,
giving 

\begin{eqnarray}
\tdel_{k}(\tau)\sim\zeta^{1}(\tau) & \sim & \epsilon^{\frac{1}{\rho-1}}.\label{eq:app:scaling_q_k}
\end{eqnarray}

\paragraph{Threshold - linear transfer function}

In Section \ref{sec:Single-inhibitory-population}, we have shown
that in one population with threshold linear transfer function ( $\nu=1$),
the excess mean input is zero at the transition (Fig. \ref{fig:Phase-diagram}).
Our numerical solutions show that in two population networks with
this transfer function, $x_{l}$ are no longer exactly zero at the
transition; nevertheless they remain small in a wide region above
the transition. Assuming $|x_{l}|\ll1$, the stability matrix can
be written as 
\begin{equation}
\tilde{M}_{kl}=\tilde{g}_{kl}^{2}H(-x_{l})\approx0.5\tilde{g}_{kl}^{2}.
\end{equation}

The mean field expression for the variance, (\ref{eq:FP_delta0})
can be written as $1=\sum_{l}\tilde{g}_{kl}^{2}\left\langle \phi^{2}\left(z+x_{l}\right)\right\rangle _{z}\approx0.5\sum_{l}\tilde{g}_{kl}^{2}$.
Hence, we find that $\sum_{l}\tilde{M}_{kl}=1+\mathcal{O}(x_{l}),\;\forall k$,
implying that the eigenvector corresponding to the largest eigenvalue
of $\mathbf{\tilde{M}}$, is uniform to leading order in $x_{k}$.
It follows that near criticality, when $|x_{k}|$ is small,  $R^{1}$is
approximately uniform, and all $q_{k}$ are nearly equal, as can be
seen in Figure \ref{fig:TwoPop_critical}.

\bibliographystyle{apsrev}
\bibliography{chaos_transition_rev2_final}

\end{document}